\documentclass[final,3p]{elsarticle}

\usepackage{amssymb,epsfig,latexsym,times,bbm,lineno}
\usepackage{amsmath}
\usepackage{breqn}
\usepackage{color}
\usepackage{mathtools, cuted}
\usepackage{dcolumn}

\usepackage{float}
\usepackage{rotating}
\usepackage{multirow}
\allowdisplaybreaks

\journal{}

\begin{document}

\title{Coupled dynamics of endemic disease transmission and gradual awareness diffusion in multiplex networks}

\author[label1]{Qingchu Wu}
\ead{wqingchu@jxnu.edu.cn}
\author[label2]{Tarik Hadzibeganovic}
\ead{tarik.hadzibeganovic@gmail.com}
\author[label3,label4]{Xiao-Pu Han}
\ead{xp@hznu.edu.cn}
\address[label1]{College of Mathematics and Information
Science, Jiangxi Normal University, Nanchang, Jiangxi
 {\rm 330022}, China}
 \address[label2]{Institute of Psychology, Faculty of Natural Sciences, University of Graz, Graz 8010, Austria}
\address[label3]{Alibaba Research Center for Complexity Sciences, Hangzhou Normal University, Hangzhou 311121, China}
\address[label4]{Institute of Information Economy and Alibaba Business College, Hangzhou Normal University, Hangzhou 311121, China}

\begin{abstract}
Understanding the interplay between human behavioral phenomena and infectious disease dynamics has been one of the central challenges of mathematical epidemiology. However, socio-cognitive processes critical for the initiation of desired behavioral responses during an outbreak have often been neglected or oversimplified in earlier models. Combining the microscopic Markov chain approach with the law of total probability, we herein institute a mathematical model describing the dynamic interplay between stage-based progression of awareness diffusion and endemic disease transmission in multiplex networks. We analytically derived the epidemic thresholds for both discrete-time and continuous-time versions of our model, and we numerically demonstrated the accuracy of our analytic arguments in capturing the time course and the steady-state of the coupled disease-awareness dynamics. We found that our model is exact for arbitrary unclustered multiplex networks, outperforming a widely adopted probability-tree-based method, both in the prediction of the time-evolution of a contagion and in the final epidemic size. Our findings show that informing the unaware individuals about the circulating disease will not be sufficient for the prevention of an outbreak unless the distributed information triggers strong awareness of infection risks with adequate protective measures, and that the immunity of highly-aware individuals can elevate the epidemic threshold, but only if the rate of transition from weak to strong awareness is sufficiently high. Our study thus reveals that awareness diffusion and other behavioral parameters can nontrivially interact when producing their effects on epidemiological dynamics of an infectious disease, suggesting that future public health measures should not ignore this complex behavioral interplay and its influence on contagion transmission in multilayered networked systems.
\end{abstract}


\date{\today}

\maketitle

\section{Introduction}   

Understanding the multiplexity of the relationship between individual behavioral responses and infectious disease trajectories in networked populations has been a grand transdisciplinary challenge \cite{Pastor-Satorras2015,Funketal2010,manfredi} and a topic of particular interest in behavioral epidemiology and biomathematics \cite{ferguson,shakeri,Huang2018,penglu,sonabend,aguiar,tangzhou,zhuliuzhang,puszynski,musa,jiading,andreagallotti,qiuzhong,qcwcsf,Wang2015}. 

Recent world-wide experience with the COVID-19 pandemic has undoubtedly demonstrated that human behavior can be critical to the transmission of contagious diseases \cite{bedsonathumbehav}, and moreover, that in the absence of vaccines or other effective pharmaceutical treatments, adoption of behavioral modifications such as social distancing \cite{aboukheydari} can often serve as the only reliable measure for controlling the outbreak dynamics \cite{perrareview}. Computer-based simulations of non-pharmaceutical responses to the early \cite{Fergimperial2020,adamsims,tirtsallis} and later \cite{samooreetal} waves of the COVID-19 outbreak have corroborated this meanwhile widespread view, even though they still remain largely limited to only specific outbreak scenarios \cite{Kupferschmidt2020}.

Given the uncertainties both in the transmission modes and in the contagiousness of the emerging infectious pathogenes such as SARS-COV-2 \cite{bertaglia}, as well as the general limitations in predictability of human behavior \cite{bellomo1,bellomo2,bellomo3}, there is a pressing need for the development of integrative mathematical models that would serve as a foundation for a general formal theory of the coupled disease-behavior spreading dynamics \cite{sahneh}. Acknowledging the significance of this multi-level relationship between human behavior and contagion spreading is necessary not only for understanding how human responses are influenced by the disease outbreaks \cite{ferguson}, but also how human behavioral reactions and the spread of awareness about the percolating pathogens can affect the propagation of the disease itself \cite{Funketal2010,bedsonathumbehav,sahneh}, resulting thereby in a coupled nonlinear dynamical system \cite{Wang2015,bianconi,weiwang} which is treatable only in limits by means of standard methods of mathematical epidemiology.

When perceiving the risks of infection in their proximate surroundings, individuals may react with preventive behaviors 
such as face-mask application \cite{Leung2020}, social distancing \cite{maharaj2012}, or if available, vaccination \cite{aguiar,samooreetal,flaig2018}, all of which can help reduce both their own and others' susceptibility to infection. Critical to these protective behavioral modifications is the diffusion of disease-related information in a given population, the emergence of elevated awareness of a circulating disease \cite{Funketal2010}, and the willingness to change and adapt one's behavior \cite{bedsonathumbehav}. Thus, having access to information about the spreading disease alone, as well as the associated initially established knowledge of potential risks, are usually not sufficient for containing the outbreak. 

Instead, elevated disease-risk awareness, its diffusion throughout the affected population, and the ultimate adoption of necessary protective behaviors are what matters most for the efficiency of non-pharmaceutical intervention (NPI) measures \cite{penglu,Fergimperial2020}. Modeling the influence of elevated awareness on outbreak evolution is therefore vital to our understanding of coupled disease-behavior spreading dynamics \cite{Funketal2010,Wang2015,bedsonathumbehav}, especially when accurate disease-related information and the associated NPI recommendations originating from reliable sources can be contaminated with noisy disinformation or misinformation coming from unreliable or less-reliable sources such as various social-media platforms \cite{penglu}.

\subsection{Models of coupled disease-behavior dynamics}

An implementation of awareness diffusion in epidemic models with structured populations typically involves methods addressing the relationship between aware and unaware network nodes that change their states dynamically and coevolve with the disease transmission \cite{Wang2015}. Following this tradition, a variety of approaches have meanwhile been developed for modeling coupled disease-awareness spreading dynamics. Among them, the mean-field approach has been widely employed for analyzing the contagion dynamics in both monoplex \cite{Wuetal2012,Zhan2018} and multiplex network topologies \cite{Funketal2010,Wang2015,bianconi}. 

However, even though it has been useful for modeling different epidemic scenarios, this method is not sufficiently general as it ignores the details of network structure and the resulting diversity of contact patterns amongst interacting individuals \cite{Pastor-Satorras2015}. As such, the method is less suitable for addressing the naturally emerging heterogeneous connectivity or the diversity of contact strengths \cite{Huetal2018} in structured populations, and their influence on both behavior and epidemic spreading. 

It has therefore been more convenient to resort to the so-called degree-based approaches, e.g. the  heterogeneous mean-field (HMF) method \cite{Liu2018} or the individual-based methods such as the discrete-time microscopic Markov-chain 
approach (MMCA) \cite{chakraleskovec2008,Gomez2010,wang2003} which has gathered enormous popularity in recent years in deriving coupled disease-information spreading models (see e.g. Table~2). Meanwhile, it has been repeatedly demonstrated that the MMCA method is indeed exact for both the susceptible-infected-susceptible (SIS) epidemics \cite{Gomez2010,Prakash2010} and for the susceptible-infected-removed (SIR) spreading process on static networks \cite{Arruda2018,Xia2019}. 

Whereas earlier studies were mostly dedicated to the analysis of coupled disease-behavior spreading processes on monoplex networks (e.g. \cite{Wuetal2012,Zhan2018}), more recent mathematical models of this coevolutionary epidemic-information interplay \cite{Wang2015,weiwang} have largely been harnessing the multiplex networks approach \cite{bianconi}. The first discrete-time model for the study of coupled disease-awareness transmission dynamics in multiplex networks \cite{Granell2013} used probability trees to derive the governing MMCA equations for the two coupled processes. The model assumed that infectious disease spreads {\it after} awareness diffusion and that recurrent-state dynamics underlies both awareness and disease transmission processes (i.e. after becoming aware, a node can go back again to the unaware state, and analogously, a once infected node can recover and go back to the susceptible state). The obtained epidemic threshold was found in this study \cite{Granell2013} to depend on the topology of the multiplex network and its interrelationship with the disease awareness. A continuous-time counterpart of this seminal epidemic-awareness model and its corresponding stability analysis were introduced 
recently in Ref. \cite{Huang2018}.

In their following study, Granell and colleagues \cite{Granell2014} addressed the influence of self-awareness and mass media on epidemics, observing that the role of self-aware individuals in contagion spreading is almost negligible, whereas the broadcast of disease-related information can critically change the fate of an outbreak. Motivated by these advances, Xia and colleagues \cite{Xia2019} developed a SIR-based epidemic model for coupled disease-awareness dynamics with mass media effects on multiplex networks, whereas other authors investigated the relevance of self-initiated reactions to disease information \cite{Kan2017} or the influence of its diversity \cite{Pan2018v1,Pan2018v2,Zang2018} on contagion transmission in multiplex networks.

In another recent study, also based on Ref. \cite{Granell2013}, Zhou and colleagues \cite{Zhouetal2019} developed an effective degree theory (EDT) of disease-awareness co-evolution on multiplex networks, and observed that the MMCA approach in fact transforms back to the HMF model as the discrete-time step approaches continuous-time limit. This study has further inspired a development of a continuous-time EDT model for concurrent three-state diffusion processes in multiplex networks \cite{ZhouIEEE2019}, that was found to outperform the previously widely adopted HMF model for continuous-time epidemic dynamics. 

Notably, in the last few years, sizable efforts have been invested in the advancement of such models with multi-state, coupled disease-behavior spreading processes on multiplex networks, including the UARU-SIS model with Maki-Thomson dynamics and different spreading time-scales \cite{silva2019}, the UAPR-SIR disease-opinion transmission with conflicting information \cite{penglu}, the sUAU-tSIS disease-awareness diffusion on simplicial complexes with time-varying spreading dynamics and memory effects \cite{wangetal2022}, the extended UAU-SIS model with time-varying self-awareness and behavioral 
responses \cite{hongetal}, or the generalized UAU-SIS dynamics with simultaneous spreading of disease and information \cite{qcwcsf}.

\subsection{Aims \& main assumptions of our model}

Notwithstanding these recent theoretical developments, most previous studies have largely ignored the fact that disease-related information is not instantaneously disseminated throughout a given population \cite{yezino}, but instead it spreads out in stages (i.e. gradually), as it passes 
on from a given individual to another via a network of communication contacts \cite{funkgilad}. This gradual progression of information spreading is further associated with awareness-transformation \cite{Guo2015}, from the very initial awareness of the presence of a threatening disease up to the highly elevated awareness of the associated disease-risks and the required behavioral modifications. In addition, diffusion of awareness of a disease during an outbreak does not need to rely on a single process, but instead it can be governed by a mixture of two distinct, competing spreading processes \cite{kanijmpc}. Ignoring these critical aspects of information propagation, earlier coupled contagion models have missed to address potential effects of stage-based progression of awareness diffusion on outbreak dynamics and its interplay with disease spreading in complex multilayered networks. A summary of awareness types that were studied in earlier epidemic models is presented in Table~3.

Stage-based progression of awareness diffusion thus implies that the perception of disease-related risks does not become instantly and uniformly distributed across the population once the information about an outbreak appears, but that instead this information propagates gradually and individuals who receive it typically undergo different stages of awareness elevation prior to becoming fully aware of the involved consequences of infection. Once this higher-level awareness has been attained, agents are more likely to modify their behavior and adequately respond to the outbreak with required self-protective measures. Moreover, from a realistic viewpoint, behavioral responses to epidemic spreading evolve typically at a slower rate than the disease-related information itself \cite{Zhan2018,Zhangh2012}, thereby naturally indicating stage-based transitions between information diffusion, the subsequent emergence of disease-related awareness, and the associated self-protective behaviors. 

Hence, it is a challenge to understand how gradual, stage-based progression of awareness diffusion would affect the epidemic spreading in coupled awareness-disease dynamic systems, and in turn, whether and how the disease propagation would influence the patterns of behavior and awareness diffusion. To address these questions, we rigorously derived a discrete-time coupled disease-awareness model based on the MMCA approach with gradual progression of awareness diffusion and recurrent-state dynamics underlying both disease and information transmission on multiplex networks, and we also developed its continuous-time version. In both the disease and communication layers of our model, we thus employed recurrent-state contagion spreading \cite{WuHadzi2018} which enabled us to study the cycling dynamics of disease and awareness transmission, as well as the associated endemic pathogen persistence \cite{wilksharkey,van2002}.

In the information layer of our model with stage-based progression of awareness propagation, we assumed that network nodes can be classified into two major types: (i) Passive awareness spreaders (also referred to as the weakly-aware $\rm W$-state nodes), who transmit the disease awareness to other individuals but themselves do not adopt any behavioral changes such as self-protective measures, and (ii) active behavior responders (referred to as the 
strongly-aware $\rm A$-state nodes), who also take part in the disease-awareness diffusion process, but in addition, they actively assess infection-related risks and then promptly respond with adequate self-protective measures to reduce their susceptibility. 

Thus, the awareness dynamics in our model has a cyclic nature of the form ``unaware $\rm (U)\rightarrow W\rightarrow A\rightarrow U$". The introduction of the $\rm W$-state nodes should thus account for those individuals who receive disease information but typically do not react instantly with protective behaviors due to e.g. their attitudes, beliefs, economic status, or societal norms. 

\subsection{Main contributions \& study overview}

The main analyses and contributions of our present study can be summarized as follows. Theoretical model predictions and stochastic simulations of the coupled UWAU-SIS awareness-disease dynamics are compared systematically on scale-free based multilayered networks. The accuracy of the analytically obtained epidemic threshold is numerically attested, and the full $\lambda-\beta^U$ phase diagram of the UWAU-SIS process is explored, comparing Monte Carlo simulations and the outcomes of our MMCA-based approach for the fraction of infectious agents in the stationary state. Furthermore, we investigated the effects of all model parameters on the epidemic threshold for a wide region of the parameter space, and we addressed the influence of network link overlaps on both the epidemic threshold and disease prevalence. 

We found that our model outperforms the previously introduced probability tree based UAU-SIS method \cite{Granell2013}, which is only a special case of our more general disease-awareness diffusion model. We further report on several novel interaction effects between behavioral model parameters that highlight the importance of stage-based progression of awareness diffusion for the future design of public health interventions. Moreover, we introduce the corresponding continuous-time version of our model, and we compare it against the discrete-time method in the determination of the epidemic threshold. 

In addition, we show how another variation of our model, based on probability trees, can also be derived, and we discuss the underlying differences between competing model versions. Overall, the predictive power of our UWAU-SIS model is illustrated via extensive simulation experiments, demonstrating that the obtained epidemic threshold condition holds for arbitrary unclustered multiplex networks.

The subsequent sections of our paper are organized as follows. In Section~2, we first introduce the basic properties of our UWAU-SIS model and the derivation of an MMCA method for the analysis of coupled disease-behavior dynamics. In Section~3, we present our results including our main analytical arguments, the obtained epidemic threshold condition, and the outcomes of our numerical experiments comparing theoretical predictions of our model with stochastic simulations on scale-free multiplex networks. We address the significance of the obtained results and sketch out some promising future study directions in Section~4, and finally, we briefly conclude in Section~5.

\section{Methods}

\subsection{Basic properties of the UWAU-SIS model}

Inspired by previous studies \cite{Granell2013,Pan2018v1,Pan2018v2,Zang2018}, we use the multiplex network approach to investigate the interplay between coupled contagion spreading and the associated awareness propagation. Multiplex network is a type of a multilayered network structure consisting of different links among the same nodes distributed across distinct network layers \cite{bianconi}. 

In our present model, the employed multiplex network structure with $N$ nodes was a two-layer duplex system consisting of one communication layer (i.e., the virtual or information layer for awareness spreading) with the adjacency matrix $(a_{ij})$, and one disease layer (i.e., the layer of physical contacts for infection spreading) with its adjacency matrix $(b_{ij})$. If nodes $i$ and $j$ are connected in the communication layer, then $a_{ij}=1$; else, $a_{ij}=0$. Similarly, if nodes $i$ and $j$ are linked in the physical contact layer, then $b_{ij}=1$, and otherwise $b_{ij}=0$.

In the physical contact layer of disease transmission, each node can modify its state in accordance with the recurrent SIS-epidemics: susceptible (S)$ \rightarrow $ infectious (I)$\rightarrow$ susceptible (S). In the virtual layer of awareness propagation, each node updates its state in accordance with the recurrent behavioral rule: unaware (U) $\rightarrow$ weakly aware ($\rm W$) $\rightarrow$ strongly aware ($\rm A$) $\rightarrow$ unaware (U). Thus, both disease and information diffusion in our model follow recurrent-state cyclic dynamics \cite{WuHadzi2018}. However, different from previous coupled contagion models, the aware nodes in our model can be classified into two categories: Weakly-aware nodes with constant susceptibility ($\rm W$), and strongly-aware nodes with reduced susceptibility ($\rm A$).

\begin{table}
\centering
\caption{List of symbols and notations used in our model}
\begin{center}
\begin{tabular}{|l|l| p{5cm}|}
\hline 
Symbol & Description \\ 
\hline
$N$ & Number of multiplex network nodes \\ 
$(a_{ij})$ & The adjacency matrix of the information layer\\ 
$(b_{ij})$ & The adjacency matrix of the disease layer\\ \hline 
$\lambda$ & The rate of information spreading from aware to unaware nodes\\ 
$\delta$ & The rate of information fading or forgetting in strongly-aware nodes\\ 
$\alpha$ & The transition rate from weakly-aware to strongly-aware nodes\\ 
$\beta^U$ & The rate of infection of unprotected nodes (a.k.a. unaware infectivity)\\ 
$\beta^A$ & The rate of infection of protected nodes (a.k.a. aware infectivity)\\ 
$\mu$ & The recovery rate of an infected network node\\ 
$\gamma$ & The rate of influence of information on disease spreading\\ \hline
$p_i^{US}(t)$ &  The
probability that at time $t$, each node $i$ is in the state $\rm US$ \\ 
$p_i^{WS}(t)$ &  The
probability that at time $t$, each node $i$ is in the state $\rm WS$ \\ 
$p_i^{AS}(t)$ &  The
probability that at time $t$, each node $i$ is in the state $\rm AS$ \\ 
$p_i^{AI}(t)$ &  The
probability that at time $t$, each node $i$ is in the state $\rm AI$ \\
$\rho^I(t)$ &  The fraction of infectious individual nodes at time $t$ \\ 
$\rho^I$ & The fraction of infectious individuals in the stationary state \\ \hline
\end{tabular}
\end{center}
\label{tab:symbol}
\end{table}

\subsection{Model states and transition rules}

For simplicity and in accordance with Ref. \cite{Granell2013}, we assume that after a node has been infected, it will automatically acquire awareness of its infection and as such, it can spread the acquired disease-related information to other individuals while its infectivity will remain unchanged. As a result of this assumption, nodes in our multiplex network model can be found in one of the four different awareness-disease states: Unaware susceptible (US) nodes, aware but unprotected susceptible ($\rm WS$) nodes (or weakly-aware susceptible), aware and protected susceptible ($\rm AS$) nodes (or strongly-aware susceptible), and finally, aware and protected but infected nodes ($\rm AI$). 

Here, {\it protected} means that after gaining strong awareness about a disease, we implicitly assume that agents have reacted with a certain self-protective behavioral response. Correspondingly, {\it unprotected} simply implies that after becoming aware of the infection, agents did not take any protective measures whatsoever.

We hereby note in passing that the choice of the above mentioned model states was intentional, primarily to enable a direct comparison of our main findings with those of previous models that were developed in the tradition of Granell et al. \cite{qcwcsf,Granell2013,Pan2018v1,Pan2018v2,wangetal2022,Guo2015,Guo2016}. Thus, as noted above, an infected node in our model automatically becomes aware of infection and then directly transmits the acquired disease-related information to other individuals. As a result, the unaware-infected (UI) state does not exist because an infected node in our model directly transitions to the strongly-aware-infected (AI) state; for related models with the included UI state, the reader is referred to Refs. \cite{Kan2017,Zhouetal2019,silva2019,velasquezrojas}. 

Importantly, the unaware-infected (UI) state should not be confused with the asymptomatic-infectious state \cite{shilong}, which has become recently popular through COVID-19 pandemic models, where it typically denotes individuals who are infected but without observable symptoms. In other words, being unaware does not mean that an individual is asymptomatic, but instead, that an individual is not conscious of potential disease-related risks, or even of the disease presence at all. 

For example, asymptomatic but infected individuals can be either unaware or aware of the actual disease risks, neither of which depends on the observability of symptoms but instead on the availability of disease-related information and its level of adoption by one's awareness. Analogously, an infected symptomatic individual can be either fully aware or unaware of the underlying disease-related risks, and consequentially may fail to adopt adequate behavioral measures such as social distancing or face-mask application.

Even much prior to the COVID-19 pandemic, modelers have used for this purpose the so-called exposed (E) compartment, which designates a latent incubation period during which a host agent is infected but not yet infectious \cite{FengWang2019}. In our case, of course, the $\rm W$-state and the associated $\rm A$- and U-states are not epidemic compartments, but instead they are awareness-related states in a separate information layer of the network that coexist and co-evolve with the disease dynamics in the contact layer. Thus, in order to model the influence of asymptomatic-infectious individuals in the context of our present model with stage-based progression of awareness diffusion, one would need an additional epidemic compartment (not the UI state), namely, the exposed E-compartment, and our current model would therefore expand to an UWAU-SEIS process (for a related UAU-SEIS process, the reader is referred to Ref. \cite{shilong}). 

We now consider the transition rules between the model states. In the virtual information layer, an unaware U node receives disease-related information from aware A neighboring nodes with rate $\lambda$. Once this disease-relevant information has been collected, the receiver node changes its state from unaware $\rm U$ to weakly-aware $\rm W$. A node in the weakly-aware $\rm W$-state can further spread the acquired epidemic information but does not undertake any protective measures. 

We further assume that a weakly-aware $\rm W$ node in our model will transition to the strongly-aware $\rm A$-state with rate $\alpha$. In addition, strongly-aware nodes can possibly lose their awareness of a disease via forgetting effects with rate $\delta$, thereby shifting from the strongly-aware state $\rm A$ to the unaware state $\rm U$. In the disease spreading layer of physical contacts, a susceptible S-node is infected through an infectious I-neighbor at infectivity rate $\beta$, and an infectious I-node can become susceptible again via recovery process at rate $\mu$. 

Considering the interaction between the virtual communication and physical contact layers, the observable infection rate in our model is linked to preventive behavior, and as such can be regarded as a kind of heterogeneous infection rate. Moreover, if the susceptible S-node enters the unaware U-state or the aware but unprotected state $\rm W$, we then have $\beta=\beta^U$, where $\beta^U$ is the initial 'unaware' infectivity \cite{Granell2013}. However, if the susceptible S-node enters the aware and protective $\rm A$ state, then $\beta=\beta^A=\gamma \beta^U$, where $\gamma \in [0,1]$ and $\beta^A$ represents the subsequent infectivity with acquired strong awareness \cite{Granell2013}. 

At $\gamma=0$, the $\rm A$ nodes are fully immune to the circulating disease \cite{Granell2013}. If, however, $\gamma=1$, the communication layer does not influence the disease layer and so the model reduces to the regular SIS epidemic process \cite{wang2003}. A summary of symbols and notations used in our current model is listed in Table \ref{tab:symbol}.

\begin{sidewaystable} 
\vspace{-12pt}
\centering
\caption{Examples of coupled disease-awareness spreading models on multiplex networks.}
\begin{center}
\begin{tabular}{llllll}\hline

\hline \\
    Authors/Year/Reference & Model Type & Description/Method/Special Features & Model States/Underlying Structure/Governing Eqs.\\ \\
   \hline
   \hline \\
  Granell et al. (2013) \cite{Granell2013} & UAU-SIS & Discrete 3-state awareness-disease model & $US$, $AS$, $AI$ \\
   &&based on MMCA and probability trees& SF and ER networks; Eq. (2)\\ \\
  Guo et al. (2015) \cite{Guo2015} & LACS-SIS & Discrete 3-state awareness-controlled disease spreading & $US$, $AS$, $AI$  \\
  &&MMCA-based with a LACS control& SF and ER networks; Eq. (2)\\ \\
  Pan \& Yan (2018) \cite{Pan2018v1} & LACS-SIS & Discrete 3-state heterogeneous LACS-SIS model & $US$, $AS$, $AI$\\
  && with 3 types of individual heterogeneity; MMCA-based & SF networks; Eq. (9) \\ \\
  Guo et al. (2016) \cite{Guo2016} & UAU-SIS & Discrete 3-state dynamic disease-awareness spreading & $US$, $AS$, $AI$  \\
  &&time-varying awareness; dynamical MMCA, probability trees &SF \& time-varying networks; Eq. (2)\\ \\
  Kan \& Zhang (2017) \cite{Kan2017} & UAU-SIS & Discrete 4-state disease-awareness model  & $SU$, $SA$, $IU$, $IA$\\
  &&with self-initiated awareness; MMCA-based & SF networks; Eqs. (6)-(10)\\ \\
  Zhou et al. (2019) \cite{Zhouetal2019} & UAP-SIR & Continuous-time 9-state disease-awareness spreading  & 
  $US$, $UI$, $UR$, $AS$, $AI$, $AR$, $PS$, $PI$, $PR$  \\
  &&based on Effective Degree Theory (EDT) and HMF & SF networks; Eqs. (1a)-(1d) and (8a)-(8i)\\ \\
  da Silva et al. (2019) \cite{silva2019} & UARU-SIS & Discrete 6-state disease-awareness spreading & $SU$, $SA$, $SR$, $IU$, $IA$, $IR$  \\
  &&with Maki-Thomson dynamics \& different time scales& SF networks; Eqs. (1)-(9) and (A5)-(A10)\\ \\
 Vel\' asquez-Rojas et al. (2020) \cite{velasquezrojas} & UAU-SIS & Discrete 4-state disease-awareness spreading & $SU$, $SA$, $IU$, $IA$  \\
  &&MF approach with varying disease-awareness spreading speeds & ER networks; Eqs. (1a)-(1d) \\ \\
  Peng et al. (2021) \cite{penglu} & UAPR-SIR & Continuous-time 12-state disease-opinion spreading  & 
  $US$, $UI$, $UR$, $AS$, $AI$, $AR$, $PS$, $PI$, $PR$, $RS$, $RI$, $RR$  \\
  &&MF-type pair-approximation for multiplex networks & CM random graphs; Eqs. (3.1)-(4.2) and (4.6)\\ \\
  Wang et al. (2022) \cite{wangetal2022} & sUAU-tSIS & Discrete-time 3-state disease-awareness spreading  & 
  $US$, $AS$, $AI$  \\
  &&MMCA-based with time-varying spreading \& memory & Simplicial complexes; Eqs. (4), (15), and (16)\\ \\
  Wu et al. (present study)  & UWAU-SIS & Discrete \& continuous-time 4-state disease-awareness model &  $US$, $WS$, $AS$, $AI$\\
  &  & based on MMCA and the law of total probability & SF networks; Eqs. (2), (6), (16), (B1) and (B10) \\ \\
  \hline
\end{tabular}
\end{center}
\end{sidewaystable} 

\subsection{The Microscopic Markov Chain Approach} 
\medskip
In our current model, we employed the MMCA method \cite{Gomez2010,wang2003} to analyze the coupled spreading dynamics. When this method is applied to our 4-state UWAU-SIS model, for $\xi\in \{U,W,A\} ,\eta\in \{S,I\}$, we have $p_i^{\xi\eta}(t)=\mathbb{P}[X_i(t)=\xi, Y_i(t)=\eta]$ which stands for the
probability that at time $t$ each individual node $i$ occupies one of the four possible model states: $\rm US, WS, AS, AI$.
In Ref. \cite{Granell2013}, probability trees were originally used to investigate the dynamical equations
of $p_i^{\xi\eta}(t)$ for the standard UAU-SIS process; for a summary of differences across related models employing the MMCA method, see Table~2.

In the present paper, we used a novel modeling framework that combines MMCA with several principles of 
probability theory. A related framework has recently been employed to derive a pairwise formulation for a 3-state UAU-SIS model with simultaneous disease-awareness spreading \cite{qcwcsf}. More generally, using the law of total probability and the property of conditional probability \cite{qcwcsf,WuHadzi2018,wuinteract}, we have

\begin{eqnarray}\nonumber
&&\mathbb{P}[X_i(t+1)=\xi, Y_i(t+1)=\eta]\\\nonumber
&=&\sum_{x,y}\mathbb{P}[X_i(t+1)=\xi, Y_i(t+1)=\eta|X_i(t)=x, Y_i(t)=y]\mathbb{P}[X_i(t)=x, Y_i(t)=y]\\\nonumber
&=&\sum_{x,y}\mathbb{P}[X_i(t+1)=\xi|Y_i(t+1)=\eta,X_i(t)=x, Y_i(t)=y]\\\nonumber
&&\quad \quad                \times\mathbb{P}[Y_i(t+1)=\eta|X_i(t)=x, Y_i(t)=y]\mathbb{P}[X_i(t)=x, Y_i(t)=y].\\\label{basic}
\end{eqnarray}
\medskip
\section{Results}

\subsection{Analytical Arguments}

Considering Eq.~(\ref{basic}), we can write a discrete-time system describing our coupled UWAU-SIS spreading dynamics on a multiplex network, which then takes the form
\begin{eqnarray}\nonumber
p_i^{US}(t+1)&=&p_i^{US}(t)r_i(t)q_i^U(t)+\delta p_i^{AS}(t) q_i^A(t) +\delta \mu p_i^{AI}(t) \\\nonumber
p_i^{WS}(t+1)&=&p_i^{US}(t)[1-r_i(t)]q_i^U(t)+(1-\alpha) p_i^{WS}(t)q_i^U(t) \\\nonumber
p_i^{AS}(t+1)&=&\alpha p_i^{WS}(t) q_i^U(t)+(1-\delta)p_i^{AS}(t)q_i^A(t)+(1-\delta)\mu p_i^{AI}(t)\\\nonumber
p_i^{AI}(t+1)&=&p_i^{US}(t)[1-q_i^U(t)]+p_i^{WS}(t)[1-q_i^U(t)]+p_i^{AS}(t)[1-q_i^A(t)]+(1-\mu)p_i^{AI}(t).\\\label{eqnn1}
\end{eqnarray}

Here, $r_i(t)$ denotes the probability that at time $t$ a node $i$ remains uninformed by any of its neighboring nodes. We further note that topologically,
the connectivity of each layer in our multiplex network is unclustered, implying independence among the states of neighboring nodes. Additionally, if a dynamical correlation of the first order is not considered, $r_i(t)$ in the UWAU-SIS process can 
be written as \cite{Gomez2010,Arruda2018}
\begin{equation}
r_i(t)=\prod_{j\in \Gamma^a_{i}}{[1-\lambda ( p_{j}^{WS}(t)+p_{j}^{AS}(t)+p_{j}^{AI}(t))]}.
\end{equation}
Here, $\Gamma^a_{i}$ represents the neighborhood of an individual node $i$ in the virtual communication layer. Furthermore, the probabilities for node $i$ of remaining uninfected via the contact with any neighboring nodes at time $t$ when $i$ was unprotected (U, W) and protected (A) are, respectively,
\begin{eqnarray}
q_i^U(t)&=&\prod_{j\in \Gamma^b_{i}}{[1-\beta^U p_{j}^{AI}(t)]}\\
q_i^A(t)&=&\prod_{j\in \Gamma^b_{i}}{[1-\beta^A p_{j}^{AI}(t)]}.
\end{eqnarray}
Here, $\Gamma^b_{i}$ stands for the neighborhood of $i$ in the contact layer of disease spreading. Similarly as in Ref. \cite{qcwcsf}, in order to elaborate on how the Eq.~(\ref{eqnn1}) results from the Eq.~(\ref{basic}), we derive as an example the corresponding equations for $p_i^{AI}(t)=\mathbb{P}[X_i(t)=A, Y_i(t)=I]$. Thus, using Eq.~(\ref{basic}), which combines conditional probability and the law of total probability, we can obtain
\begin{eqnarray}\nonumber
&&\mathbb{P}[X_i(t+1)=A, Y_i(t+1)=I]\\\nonumber
&=&\sum_{x,y}\mathbb{P}[X_i(t+1)=A, Y_i(t+1)=I|X_i(t)=x, Y_i(t)=y]\mathbb{P}[X_i(t)=x, Y_i(t)=y]\\\nonumber
&=&\sum_{x,y}\mathbb{P}[X_i(t+1)=A|Y_i(t+1)=I,X_i(t)=x, Y_i(t)=y]\\\nonumber
&&\quad \quad                \times\mathbb{P}[Y_i(t+1)=I|X_i(t)=x, Y_i(t)=y]\mathbb{P}[X_i(t)=x, Y_i(t)=y]
\end{eqnarray}
\begin{eqnarray}\nonumber
&=&\mathbb{P}[X_i(t+1)=A|Y_i(t+1)=I,X_i(t)=U, Y_i(t)=S]\mathbb{P}[Y_i(t+1)=I|X_i(t)=U, Y_i(t)=S]\\\nonumber
&&\quad \quad   \quad \quad                \times\mathbb{P}[X_i(t)=U, Y_i(t)=S]\\\nonumber
&&+\mathbb{P}[X_i(t+1)=A|Y_i(t+1)=I,X_i(t)=W, Y_i(t)=S]\mathbb{P}[Y_i(t+1)=I|X_i(t)=W, Y_i(t)=S]\\\nonumber
&&\quad \quad   \quad \quad                \times\mathbb{P}[X_i(t)=W, Y_i(t)=S]\\\nonumber
&&+\mathbb{P}[X_i(t+1)=A|Y_i(t+1)=I,X_i(t)=A, Y_i(t)=S]\mathbb{P}[Y_i(t+1)=I|X_i(t)=A, Y_i(t)=S]\\\nonumber
&&\quad \quad   \quad \quad                \times\mathbb{P}[X_i(t)=A, Y_i(t)=S]\\\nonumber
&&+\mathbb{P}[X_i(t+1)=A|Y_i(t+1)=I,X_i(t)=A, Y_i(t)=I]\mathbb{P}[Y_i(t+1)=I|X_i(t)=A, Y_i(t)=I]\\\nonumber
&&\quad \quad   \quad \quad                \times\mathbb{P}[X_i(t)=A, Y_i(t)=I].\\
\label{derivation}
\end{eqnarray}
As we mentioned previously in our Section~2.2, following its infection in the contact layer, a given node automatically becomes aware of it and accordingly modifies its state in the communication layer to the strongly-aware-infected state $\rm AI$. For all $x,y$, we therefore have $\mathbb{P}[X_i(t+1)=A|Y_i(t+1)=I,X_i(t)=x, Y_i(t)=y]=1$. When $X_i(t)=U$, the unaware but susceptible nodes are infected with the probability $1-q_i^U(t)$ \cite{qcwcsf}. Thus,
$$
\mathbb{P}[Y_i(t+1)=I|X_i(t)=U, Y_i(t)=S]=1-q_i^U(t).
$$
Also, when $X_i(t)=A$, the aware but susceptible nodes will be infected at a rate $1-q_i^A(t)$ and we thus have
$$
\mathbb{P}[Y_i(t+1)=I|X_i(t)=A, Y_i(t)=S]=1-q_i^A(t).
$$

Finally, one can easily show that
$$
\mathbb{P}[Y_i(t+1)=I|X_i(t)=A, Y_i(t)=I]=1-\mu.
$$
Therefore, upon substituting the above equations into Eq.~(\ref{basic}) we can immediately derive the equations for $p_i^{AI}(t)$. Related dynamical equations for other variables can also be obtained in a similar fashion. From this presented analysis, one can conclude that our described coupled epidemic model holds approximately for a multiplex network whose individual layers are not clustered. 

We would like to stress here that our MMCA-based UWAU-SIS model is originally derived by using the total probability formula rather than the probability tree based method, which has been used widely in earlier models of coupled epidemic-behavior dynamics. Thus, as in Ref. \cite{qcwcsf}, in our main discrete-time model formulation, we assume that at each time step, awareness and disease processes always spread concurrently rather than sequentially alternating between each other, as e.g. is the case in Ref. \cite{Granell2013}, where disease transmission occurs after information diffusion. However, by using the previously introduced method \cite{Granell2013}, we show in Appendix A that the corresponding discrete-time model version (\ref{eqn1}) with sequential spreading can also be derived for our 4-state UWAU-SIS process with coupled endemic disease transmission and gradual awareness diffusion. 

Thus, besides our main model of the UWAU-SIS process with concurrently coupled spreading dynamics, we demonstrate in Appendix A that the corresponding model version can be established by using a different, probability tree based method, in which disease spreads {\it after} information diffusion, as in Ref. \cite{Granell2013}. In addition, besides this main formulation for the discrete-time UWAU-SIS model, we derived its corresponding continuous-time version in Appendix B. It can be observed that there exist conspicuous differences between the mentioned model versions. Related model variations for the discrete-time 3-state UAU-SIS process on multiplex networks \cite{Granell2013} with simultaneously coupled disease-information spreading or with sequential transmission dynamics were studied in Ref. \cite{qcwcsf}.

\begin{sidewaystable} 
\centering
\caption{Types of awareness states in coupled disease-awareness spreading models.}
\begin{center}
\begin{tabular}{llllll}\hline

\hline \\
    Authors/Year/Reference & Types of awareness states & Description/Characteristics\\ \\
   \hline
   \hline \\
  Guo et al. (2015) \cite{Guo2015} & local awareness & attained when an unaware node is infected or the ratio between\\
   && a node's aware neighbors and its degree reaches a critical point\\ \\
  Wu et al. (2012) \cite{Wuetal2012} & local awareness & decreases the likelihood of an outbreak  \\
  &global awareness&does not decrease the likelihood of an outbreak \\ 
  &contact awareness & decreases the likelihood of an epidemic outbreak\\ \\
  Sun et al. (2018) \cite{Sunetal2018} & adaptive behavior & responsive behavior due to adaptively modified disease-related awareness \\
  &behavioral info-transmission& the propagation of disease-related awareness itself \\ \\
  Kan \& Zhang (2017) \cite{Kan2017} & self-initiated awareness & attained via infected neighbors or mass-media effects  \\
  &informed awareness&attained when informed by other aware neighbors \\ \\
  da Silva et al. (2019) \cite{silva2019} & stifler awareness & a node $i$ is informed/aware but does not transmit disease-awareness to $j$ others  \\
  &awareness&$i$ passes information to $j$ others; if $j$ nodes are stiflers or aware, $i$ becomes a stifler \\ 
  &self-awareness&with probability $\kappa$, an infected-unaware node $i$ can become aware of its own condition \\ \\
  Hong et al. (2022) \cite{hongetal} & time-varying self-awareness & self-initiated awareness changes over time  \\
  &time-varying behavioral response&behavioral responses to outbreak change over time \\ \\
  Wang et al. (2022) \cite{wangetal2022} & simplicial awareness & awareness transmission over 2-simplicial complexes \\
  & &awareness diffusion with synergistic reinforcement effect \\ \\
  Wu et al. (present study)  & weak awareness & agent transmits disease-awareness to others without taking any protective measures  \\
  &strong awareness  &  agent transmits disease-awareness to others and takes adequate protective measures  \\ \\
  \hline
\end{tabular}
\end{center}
\end{sidewaystable} 

\subsection{The epidemic threshold}

The epidemic threshold can be understood as a tipping point above which the disease breaks out and then persists in the infected population. As such, the epidemic threshold represents a critical quantity in any epidemiological modeling research. In order to determine the epidemic threshold of the system (\ref{eqnn1}), we write a linear expression of the model around the fixed points \cite{Granell2013} $p_i^{US}(t)=p_i^{US},p_i^{WS}(t)=p_i^{WS},p_i^{AS}(t)=p_i^{AS},p_i^{AI}(t)=0$. Firstly, from Eq.~(\ref{eqnn1}), one finds that the equilibrium states satisfy the following relations:
\begin{eqnarray}\nonumber
p_i^{US}&=&p_i^{US}r_i+\delta p_i^{AS} \\\nonumber
p_i^{WS}&=&p_i^{US}(1-r_i)+(1-\alpha) p_i^{WS} \\\nonumber
p_i^{AS}&=&\alpha p_i^{WS} +(1-\delta)p_i^{AS}.\\
\label{relation1}
\end{eqnarray}
Hence,
\begin{eqnarray}\label{relation2}
p_i^{WS}=\frac{1-r_i}{\alpha}p_i^{US},\quad\quad p_i^{AS}&=&\frac{1-r_i}{\delta}p_i^{US}
\end{eqnarray}
with
\begin{eqnarray}\label{a1}
r_i= \prod_{j\in \Gamma^a_{i}}{[1-\lambda (p_{j}^{WS}+p_{j}^{AS})]}.
\end{eqnarray}
We then have \cite{qcwcsf}
\begin{eqnarray}\label{a2}
q_i^U=1-\beta^U\sum_{j\in \Gamma^b_{i}}{p_{j}^{AI}},\quad \quad q_i^A=1-\beta^A\sum_{j\in \Gamma^b_{i}}{p_{j}^{AI}}.
\end{eqnarray}
Note that the fourth equation of (\ref{eqnn1}) can also be further rewritten as
\begin{eqnarray}\label{eqn2}
p_i^{AI}(t+1)&=&p_i^{AI}(t)(1-\mu)+(1-q_i^A(t)) p_i^{AS}(t)
+(1-q_i^U(t))\left[p_i^{US}(t)+p_i^{WS}(t)\right].
\end{eqnarray}
By using Eqs.(\ref{relation1})-(\ref{a2}), the linear system of Eq.~(\ref{eqn2}) is then given by
\begin{eqnarray}\label{lin1}
p_i^{AI}(t+1)&=&p_i^{AI}(t)(1-\mu)+\left[\beta^Ap_i^{AS}+\beta^U(p_i^{US}+p_i^{WS})\right]\sum_{j\in \Gamma^b_{i}}{p_{j}^{AI}(t)}.
\end{eqnarray}

\noindent Next, the $N\times N$ system matrix $R$ is defined as \cite{qcwcsf}
$$
R=(1-\mu) I+\beta^U H,
$$
where $I$ represents a unit matrix and $H$ needs to satisfy
\begin{eqnarray}\label{jacobin}
h_{ij}=\left(\gamma p_i^{AS}+p_i^{US}+p_i^{WS}\right)b_{ij}=\left[\left(\frac{\gamma}{\delta}+\frac{1}{\alpha}\right)(1-r_i)+1\right]p_i^{US}b_{ij},
\end{eqnarray}
where
\begin{eqnarray}
r_i= \prod_{j\in \Gamma^a_{i}}{\left[1-\lambda \left(p_{j}^{WS}+p_{j}^{AS}\right)\right]}=\prod_{j\in \Gamma^a_{i}}{\left[1-\lambda \left(1-p_{j}^{US}\right)\right]},
\end{eqnarray}
and
$$
\left[\left(\frac{1}{\delta}+\frac{1}{\alpha}\right)(1-r_i)+1\right]p_i^{US}=1.
$$
Then, by considering the above matrix $R$, the Eq.(12) can be described by the following linear system \cite{qcwcsf}
\begin{eqnarray}\label{lin2}
p_i^{AI}(t+1)&=&\sum_{j}R_{ij}p_j^{AI}(t).
\end{eqnarray}

This system (\ref{lin2}) will be stable if all the eigenvalues of its matrix $R$ are smaller than one, which in accordance with the spectral radius $\rho(R)$ definition is simply $\rho(R)<1$ \cite{qcwcsf}. Since the $N\times N$ system matrix $R$ is actually a Metzler matrix, its largest eigenvalue is $ \Lambda_{\rm max}(R)$ which is a real number \cite{qcwcsf,Granell2013}. Once the matrix $H$ is irreducible we obtain via the Perron-Frobenius
theorem that $\rho(R)= \Lambda_{\rm max}(R)$ \cite{qcwcsf}.
This further indicates that the critical threshold for the disease outbreak is \cite{Granell2013}
 \begin{equation}\label{threshold}
 \beta_c^U=\frac{\mu}{\Lambda_{\rm max}(H)}.
 \end{equation}
Furthermore, from Eq.~(\ref{jacobin}), one can show that the matrix $H$ can also be rewritten as
\begin{eqnarray}\label{jacobinv1}
h_{ij}=\left[1-\frac{\alpha(1-\gamma)}{\alpha+\delta}\left(1-p_i^{US}\right)\right]b_{ij},
\end{eqnarray}
whereby $p_i^{US}$ obeys the condition
$$
\left[\left(\frac{1}{\delta}+\frac{1}{\alpha}\right)\left(1-\prod_{j\in \Gamma^a_{i}}{\left[1-\lambda \left(1-p_{j}^{US}\right)\right]}\right)+1\right]p_i^{US}=1.
$$
In particular, when $\alpha=+\infty$, the epidemic threshold becomes theoretically just the same as that derived for the coupled UAU-SIS process in Ref. \cite{Granell2013}.

\subsection{Simulation Results}

\subsubsection{Simulation experiments for the UWAU-SIS dynamics and the epidemic threshold}

\begin{figure}[t]
  \centering
  \hspace*{-1.3em}
\begin{minipage}[c]{.5\textwidth}
\centering \scalebox{.3}{\includegraphics{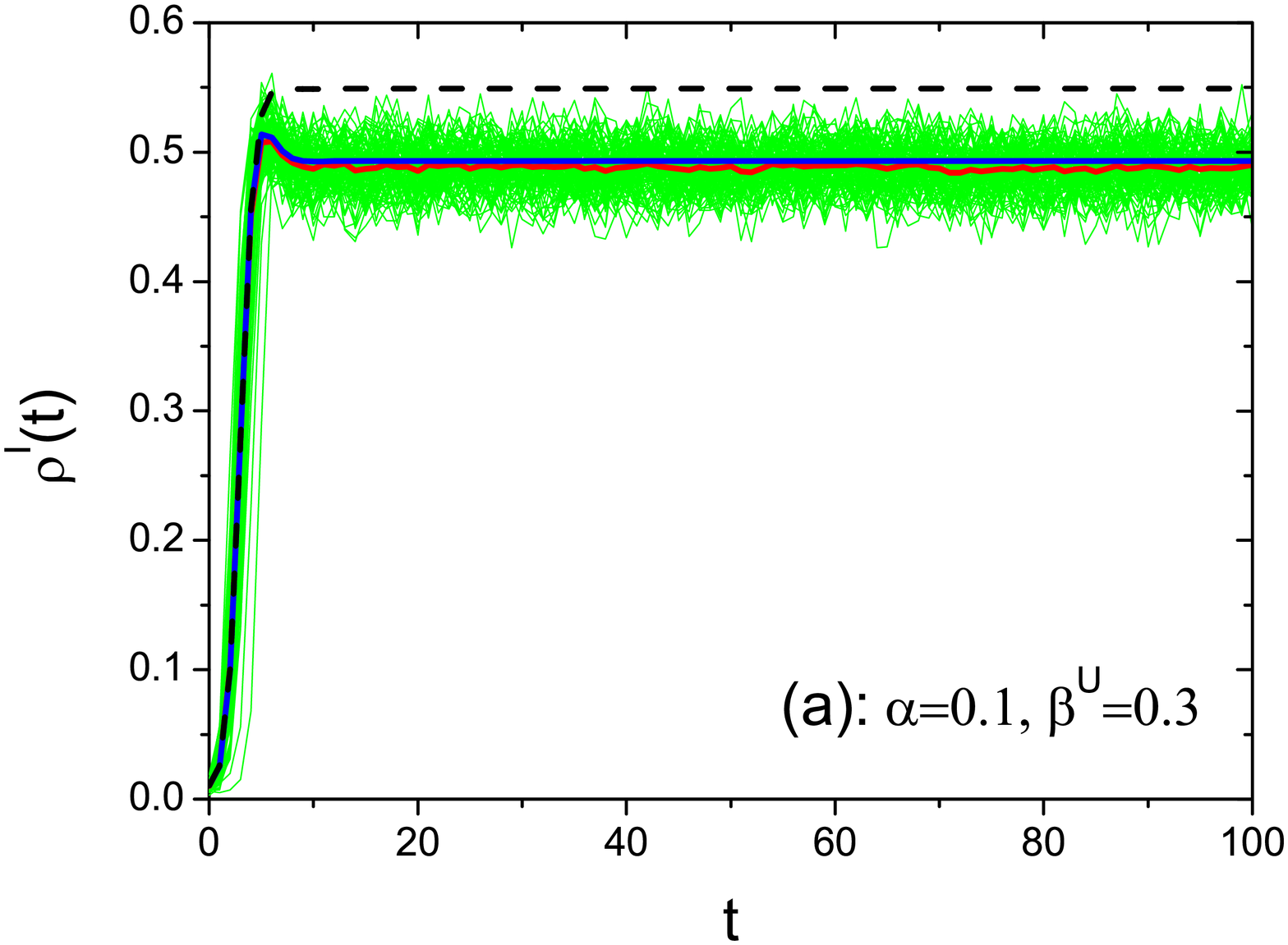}}
\end{minipage}%
\hspace{-23pt}
\begin{minipage}[c]{.5\textwidth}
\centering \scalebox{.3}{\includegraphics{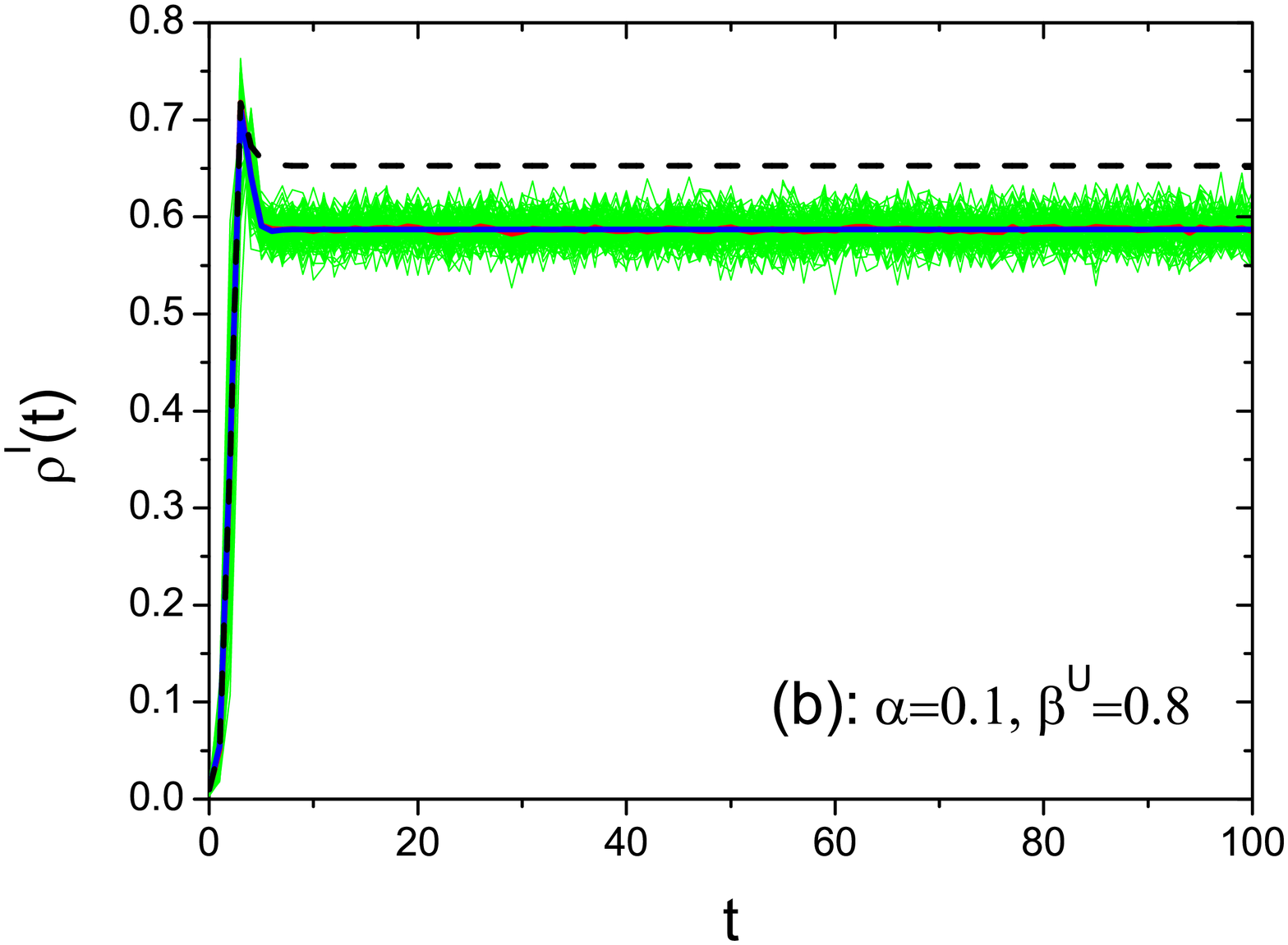}}
\end{minipage}\\[5pt] \vspace{-10pt}
\hspace*{-1.3em}
\begin{minipage}[c]{.5\textwidth}
\centering \scalebox{.3}{\includegraphics{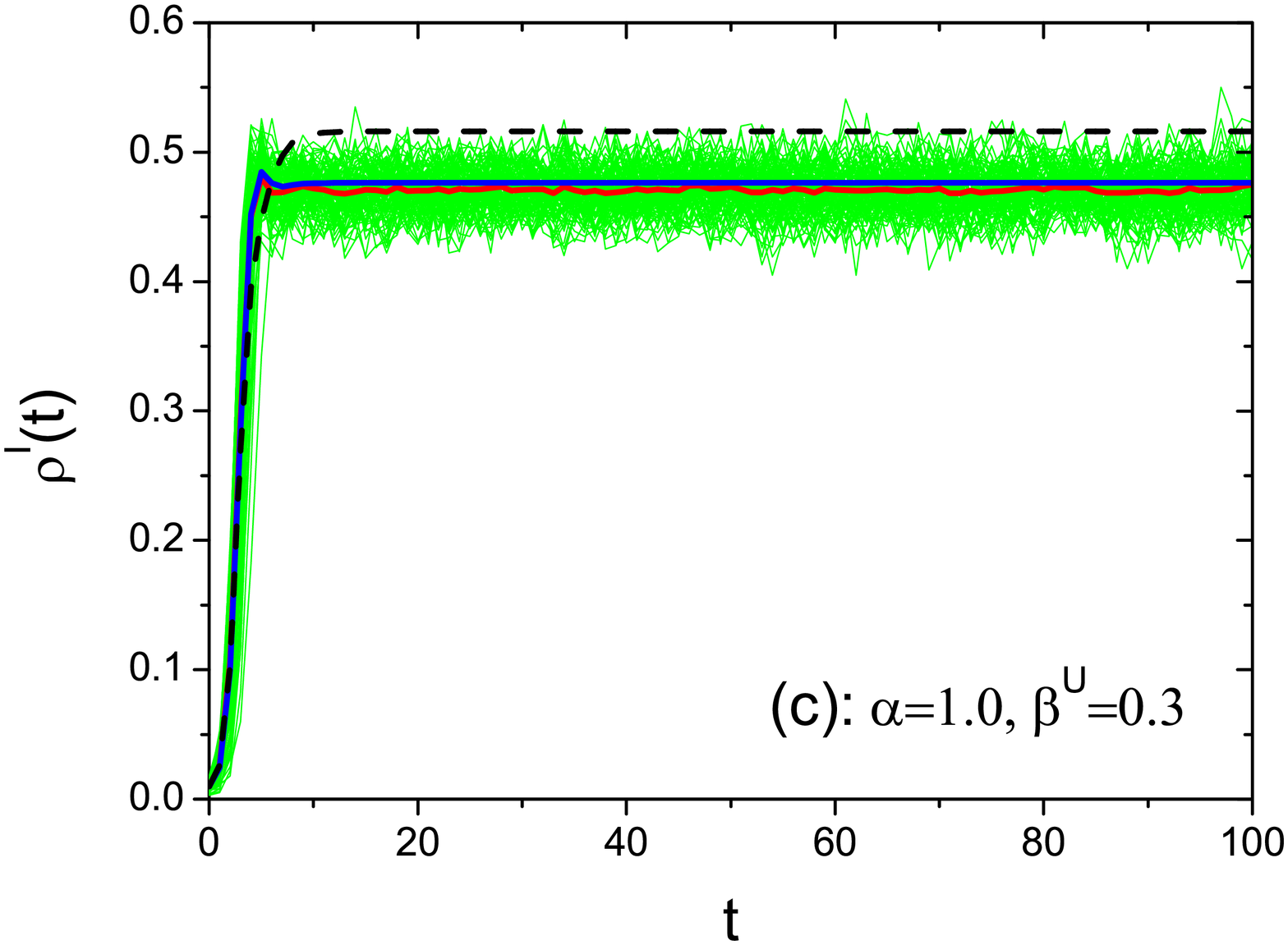}}
\end{minipage}%
\hspace{-23pt}
\begin{minipage}[c]{.5\textwidth}
\centering \scalebox{.3}{\includegraphics{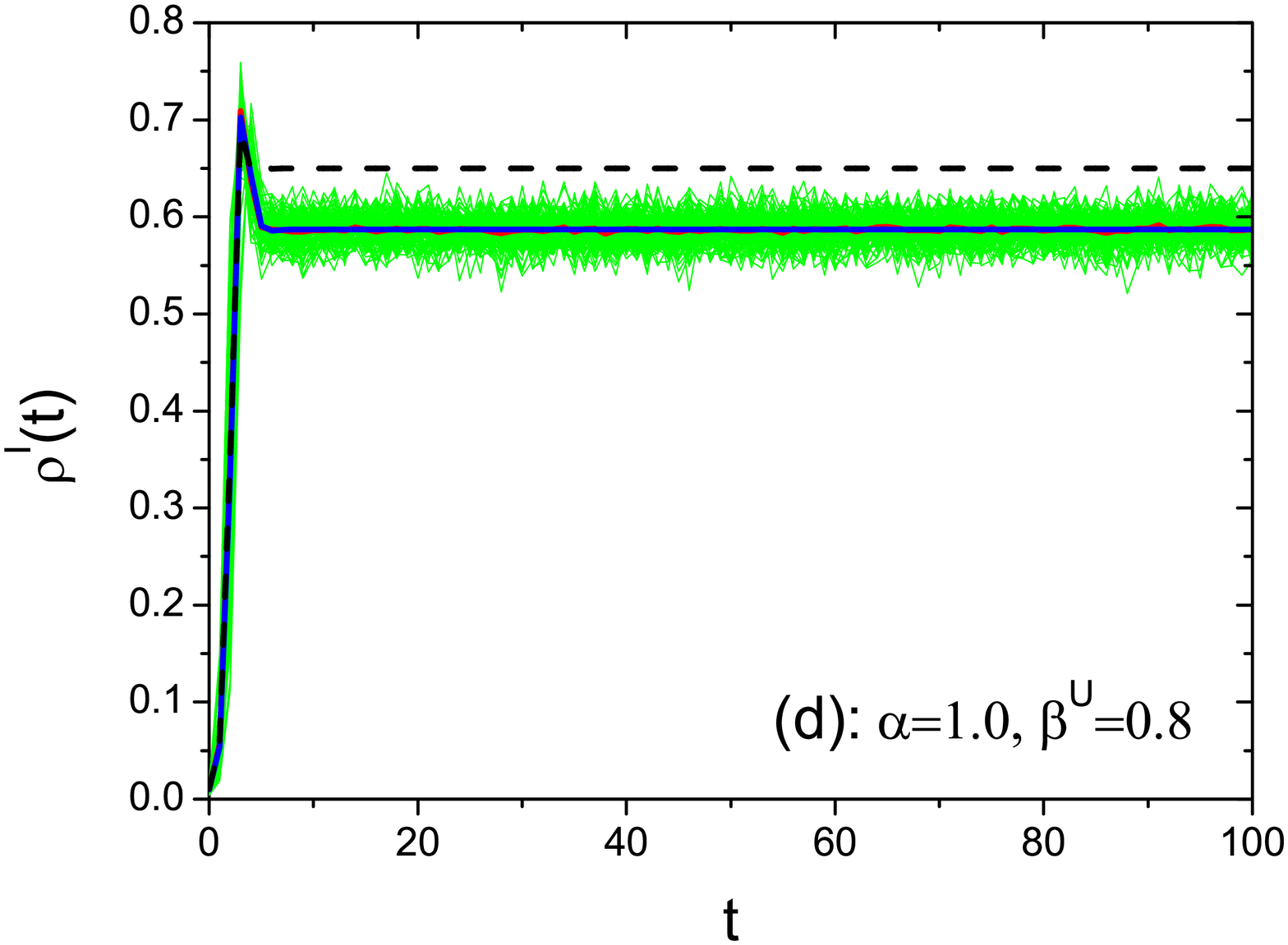}}
\end{minipage}\\[5pt]
\vspace{-10pt}
\caption{ (Color online) The time-evolution of infection density $\rho^I(t)$ in coupled disease-awareness spreading models at a fixed rate of transition from weakly-aware to strongly-aware nodes $\alpha$ and at a fixed infectivity rate $\beta^U$.  
The red curves in all panels represent the ensemble averages of 100 independent stochastic simulation runs (depicted by green lines) that are further contrasted against the numerical outcomes of two theoretical models, including our new UWAU-SIS model (\ref{eqnn1}) predictions (blue line) and the predictions obtained with the corresponding probability tree based model (\ref{eqn1}) (dashed black line) in a scale-free multiplex network with the parameter values $q=0.2$, $k_{0} = 3$, $\Delta = 2.5$, and $N = 1000$. The model parameter values employed in our stochastic simulations were $\gamma=0$, $\delta=0.6$, $\lambda=0.15$, $\mu=0.4$, and in the initial condition there were always $20$ infected/aware nodes. Panels (a)-(d) summarize the results for different combinations of values of $\alpha,\beta^U$ parameters.}
\end{figure}

In the previous section, we obtained the conditions for an epidemic
outbreak (\ref{threshold}). From this derivation, one can see that all considered dynamical parameters in our model ($\alpha,\beta^U,\gamma,\mu, \lambda,\delta$) can influence the
epidemic threshold. In this section of the article, we will assess numerically both the theoretical model (\ref{eqnn1}) and the obtained epidemic threshold condition (\ref{threshold}) by means of extensive computer simulations conducted on multiplex scale-free (SF) networks.

Our multiplex network is overlapped by two SF networks with the node degree following a power-law distribution $P(k)\sim k^{-\Delta}$ ($k_0\leq k\leq k_c$). The SF network is 
constructed by using the standard configuration model \cite{Newman2001} with the parameters $k_0=3$ and $k_c=N^{\frac{1}{\Delta-1}}$. All simulations started with 20 initially aware and infectious seeds, and then the governing rules of the UWAU-SIS spreading process with parallel updating were iterated until finally a convergence of the system to a steady-state was observed. Furthermore, random fluctuations emerging from varying initial conditions were minimized by obtaining 100 independent simulation runs, each containing different initially infectious nodes.

\begin{figure}[t]
  \centering
\begin{minipage}[c]{.5\textwidth}
\hspace*{-3.3em}
\centering \scalebox{.3}{\includegraphics[24,7][677,563]{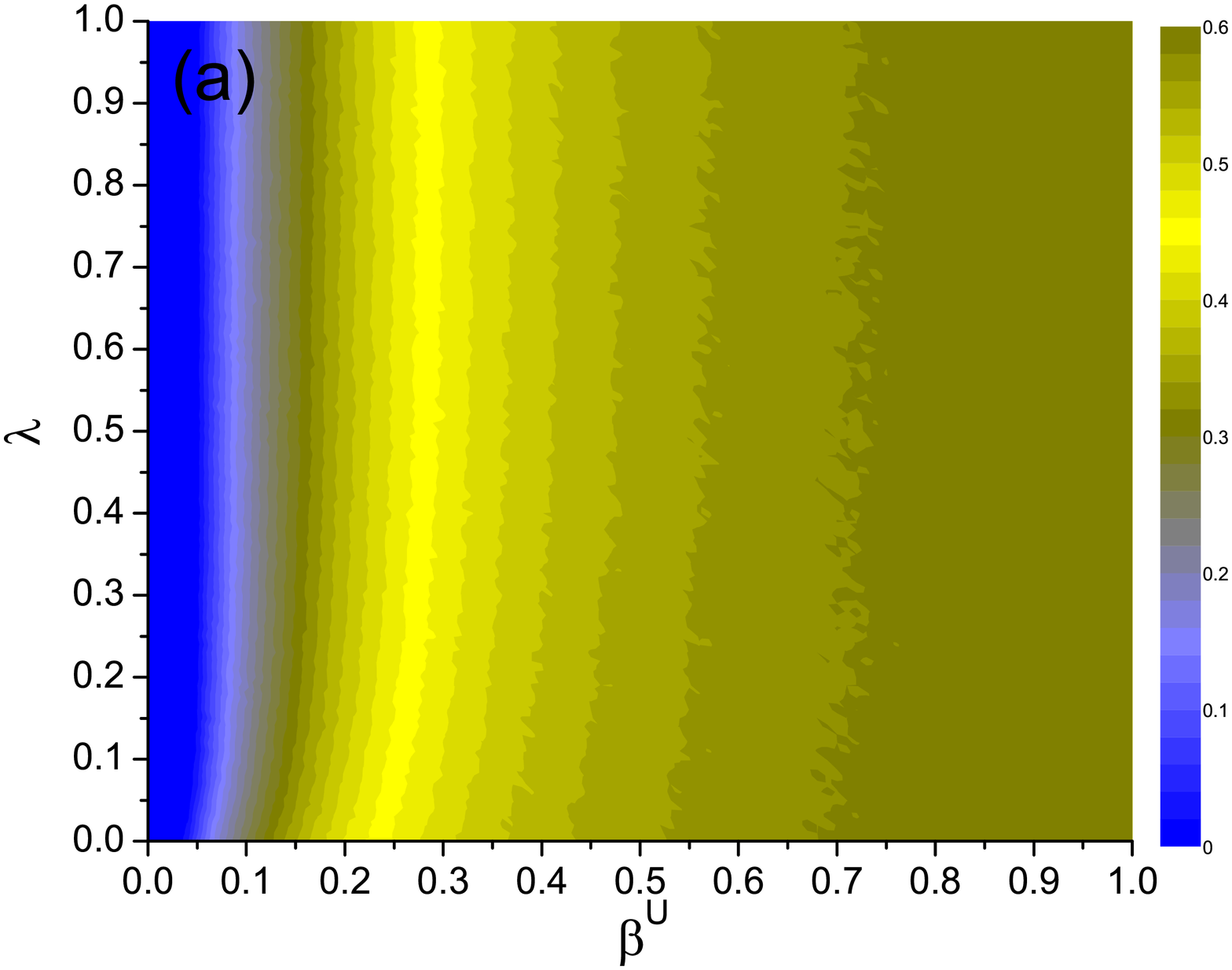}}
\end{minipage}%
\begin{minipage}[c]{.5\textwidth}
\hspace*{-3.3em}
\centering \scalebox{.3}{\includegraphics[24,7][677,563]{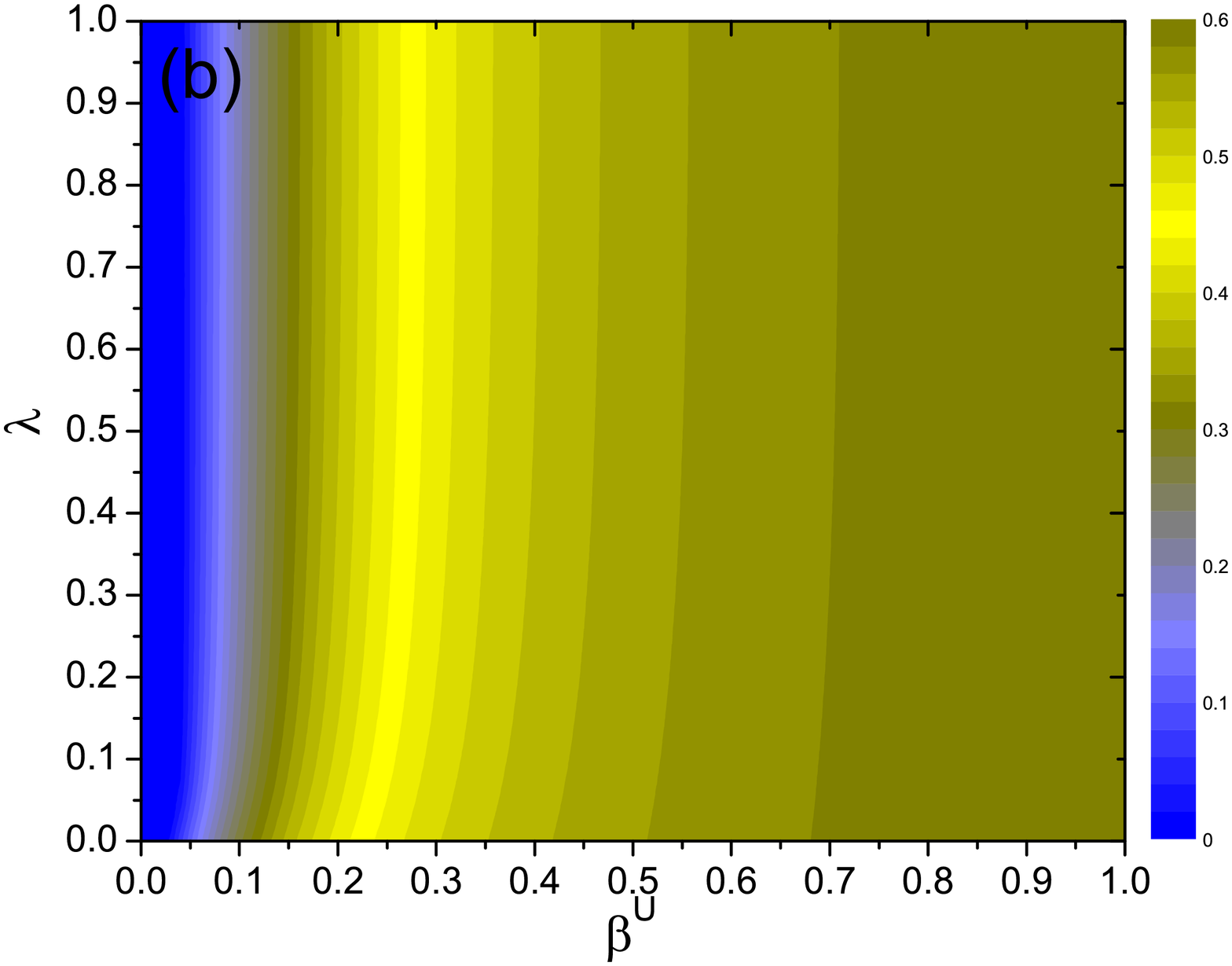}}
\end{minipage}\\[5pt]
\vspace{-5pt}
\caption{(Color online) Full phase diagrams $\lambda-\beta^U$ for the same multiplex as described in Figure~1, contrasting Monte Carlo simulations 
and the MMCA approach for the fraction $\rho^I$ of infected individual agents in the stationary state (colors represent the fractions of infected individuals). Panel (a) shows the stochastic simulation results and panel (b) represents the numerical simulations of the MMCA model. In all simulations, our selected model parameter values were $\delta=0.6$, $\mu=0.4$, $\gamma=0$,$\alpha=0.5$ and $q=0.2$.}
\end{figure}

We generated a set of multiplex network structures by means of edge 
rewiring procedure to address the link overlap effects \cite{yang2016,Zhu2019}. More specifically, we initially generated an SF network to represent both the virtual communication and the disease spreading layers. Then, in the communication network layer, we successively selected one link that was present in both network layers and one more link without a common endpoint, and we then swapped them with a rate $q$ by modifying their endpoints while avoiding the emergence of new overlapping links or multi-link formations in the communication layer. Finally, as soon 
as we could no longer select any further links, the final virtual communication layer was regarded as the disease layer. 

When $q=0$, the two SF network layers are characterized by their complete overlap. In the extreme case of $q=1$, no single link overlap can be observed across network layers. Unless otherwise specified, our stochastic simulations started with 20 initially infectious/aware seeds, and for a better comparability with earlier studies \cite{qcwcsf,Granell2013,Kan2017}, the network size of our simulated systems was kept to $N=10^3$, the coefficient $\gamma=0$, and the degree exponent was set to $\Delta=2.5$. It is worth noting here that selecting $\gamma=0$ is usually better for highlighting the dynamic interplay between information diffusion and disease spreading processes \cite{Granell2013}. Otherwise, as $\gamma$ approaches unity, the information diffusion layer has less influence on the disease spreading layer, so that at $\gamma=1$ the model ultimately reduces to the regular SIS process \cite{wang2003} in which information propagation does not exhibit any influence on the disease dynamics whatsoever. Other model parameters were also set to their baseline values in accordance with most earlier studies, and were further systematically varied in Section~3.3.2.

In the four panels of Figure~1, we show the infection density $\rho^I(t)$ vs. time $t$, by contrasting the outcomes of stochastic simulation runs with our UWAU-SIS model predictions and the theoretical predictions of the probability tree based method. Here, the infection density $\rho^I(t)$ is defined as \cite{Granell2013}
$$
\rho^I(t)=\frac{1}{N}\sum_{i=1}^Np_i^{AI}(t).
$$
Green lines in our Figure~1 represent the outcomes of individual stochastic simulation runs and their corresponding mean values are depicted as red curves. Theoretical model outcomes are represented by blue lines (\ref{eqnn1}) and dashed black lines (\ref{eqn1}), respectively. From Figure~1, we can observe that nearly throughout the whole evolution of the spreading process, the predictions of our proposed model (\ref{eqnn1}) closely fit the simulation outcomes, while simultaneously revealing a rather large discrepancy with the predictions of the probability tree based model (\ref{eqn1}). This indicates that our modified MMCA method is indeed useful for modeling the coupled disease-information dynamics with gradual awareness diffusion under the multiplex network framework. 

Interestingly, we can also see in Figure~1 that initially, but only at a sufficiently high infection rate such as $\beta^U=0.8$, the density of infected nodes at first peaks to a relatively large value but then rapidly drops to an endemic steady-state. Further comparisons among theoretical model predictions and stochastic simulation experiments with respect to the final epidemic size can be found in Figure~5. 

We also explored the full $\lambda-\beta^U$ phase diagrams of the UWAU-SIS dynamics in Figure~2, comparing Monte Carlo simulation 
outcomes (in Figure~2(a)) and the simulations of the MMCA approach (Figure~2(b)) for the fraction $\rho^I$ of infectious individual agents in the stationary state (figure colors represent different fractions of infected individuals). As can be observed from Figure~2, there was a highly satisfactory agreement between the two methods for the full phase space.

Next, we illustrated the accuracy of the derived epidemic threshold (\ref{threshold}). To this effect, we varied the infection rate $\beta$ to study its influence on the changes in the epidemic prevalence, which represents the population fraction infected at the steady state of the theoretical model (\ref{eqnn1}). In Figure~3, we plot the resulting epidemic prevalence as a function of the contagion rate $\beta\in [0, 1]$ for $\alpha=0$ and $\alpha=1$, and additionally, the corresponding epidemic threshold $\beta_c^U$ is numerically computed by using the formula (\ref{threshold}). We can see that the epidemic threshold obtained by our linear stability analysis is indeed accurate for the present model.
\begin{figure}[htb]
  \centering
\scalebox{.32}{\includegraphics[21,10][776,560]{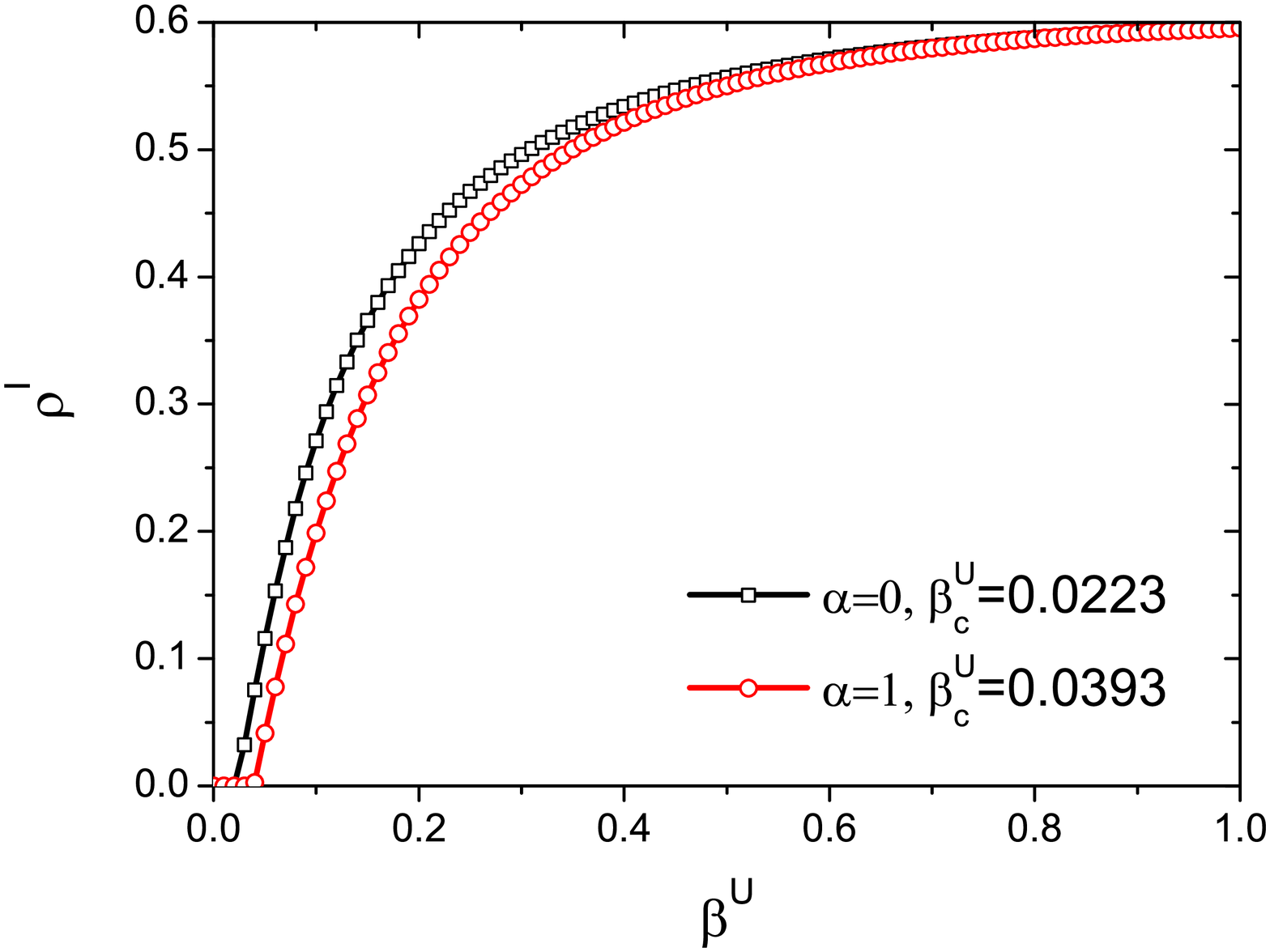}}
\caption{(Color online) The density of infectious nodes $\rho^I$ in the stable state as a function of $\beta^U$ for two different values of $\alpha$ in a multiplex network with $q=0.2,\Delta= 2.5$, $k_{0}= 3$, and $N=1000$. Our other model parameter values were: $\lambda=0.15,\delta=0.6,\gamma=0,\mu=0.4.$ The displayed simulation results represent the performance of the MMCA model (\ref{eqnn1}), and the epidemic threshold is computed with the formula (\ref{threshold}).
}\vspace{0.5cm}
\end{figure}

\subsubsection{Influence of the model parameters on epidemic spreading}

In Figure~3 we can further observe that the parameter $\alpha$ affects the epidemic threshold such that $\beta_c^U$ increases with $\alpha$. The same effect is also confirmed in Figure~4(a). Importantly, unlike at large values of $\alpha$, when $\alpha$ is small ($\alpha<0.2$), the information transmission rate $\lambda$ will have little or no influence on the epidemic threshold (Figure~4(a)). Thus, there is an interaction effect between $\alpha$ and $\lambda$, indicating that the rate at which the disease-related information is transmitted from aware to unaware individuals influences an epidemic outbreak but only if $\alpha$ is sufficiently large, i.e. if the rate at which weakly aware W-individuals transition to strongly aware A-individuals is adequately high.

Next, after identifying the importance of $\alpha$ in our model, we systematically check the influence of other model parameters on the epidemic threshold under distinct values of $\alpha$. In Figure~4(b), we see that the epidemic threshold rises with $\lambda$, and even more so with larger $\alpha$. Interestingly, from these curves in Figure~4(b), we can observe a
certain critical value $\lambda_c$: When $\lambda<\lambda_c$, the contagion threshold does not depend on information spreading; $\lambda_c$ is called the {\it metacritical point} \cite{Granell2013}. It can be explained by deriving the specific expression of the epidemic threshold for a regular multiplex network as a special case. 

\begin{figure}[t]
  \centering
\begin{minipage}[c]{.5\textwidth}
\centering \scalebox{.3}{\includegraphics{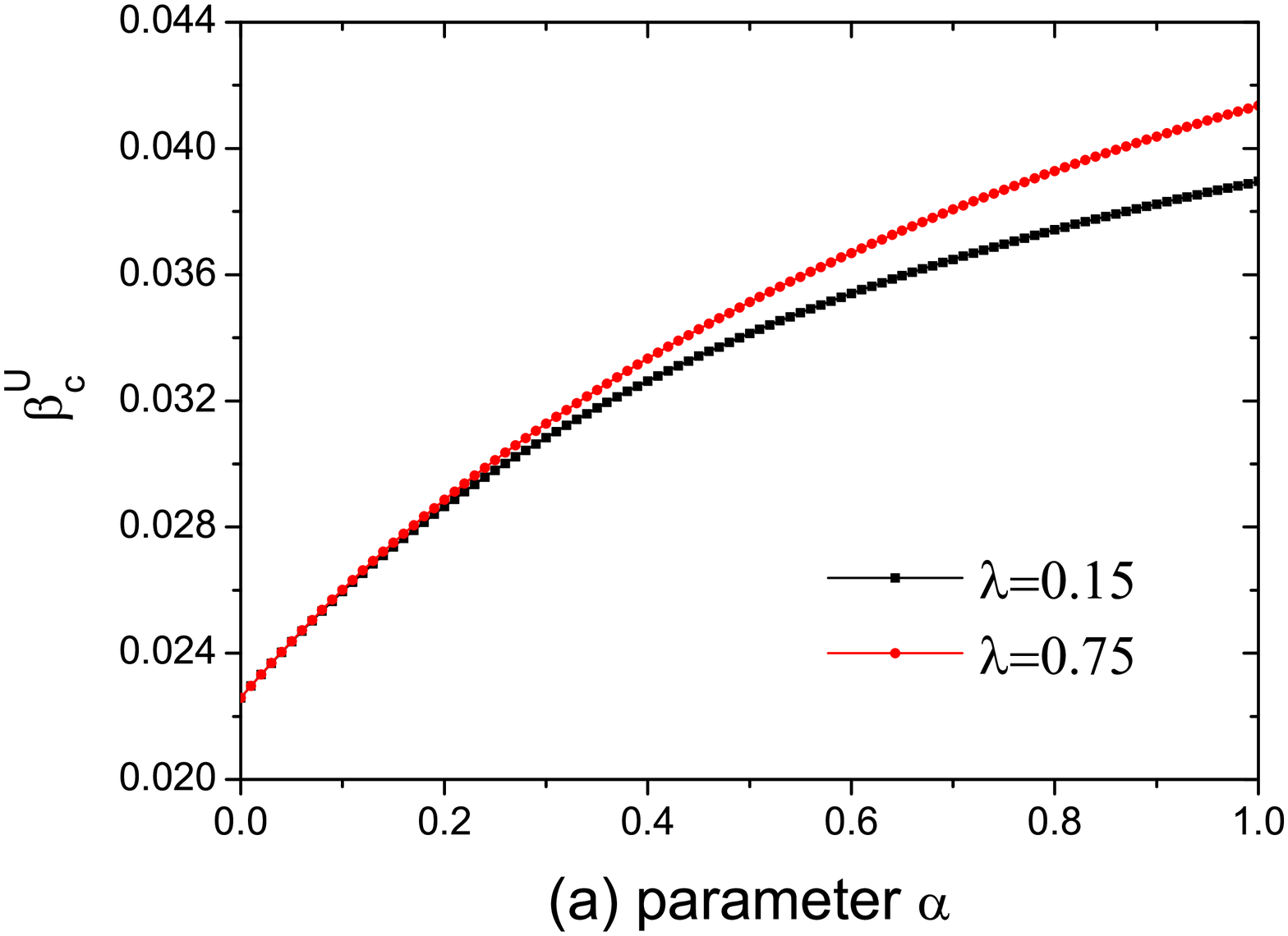}}
\end{minipage}%
\hspace{-14pt}
\begin{minipage}[c]{.5\textwidth}
\centering \scalebox{.3}{\includegraphics{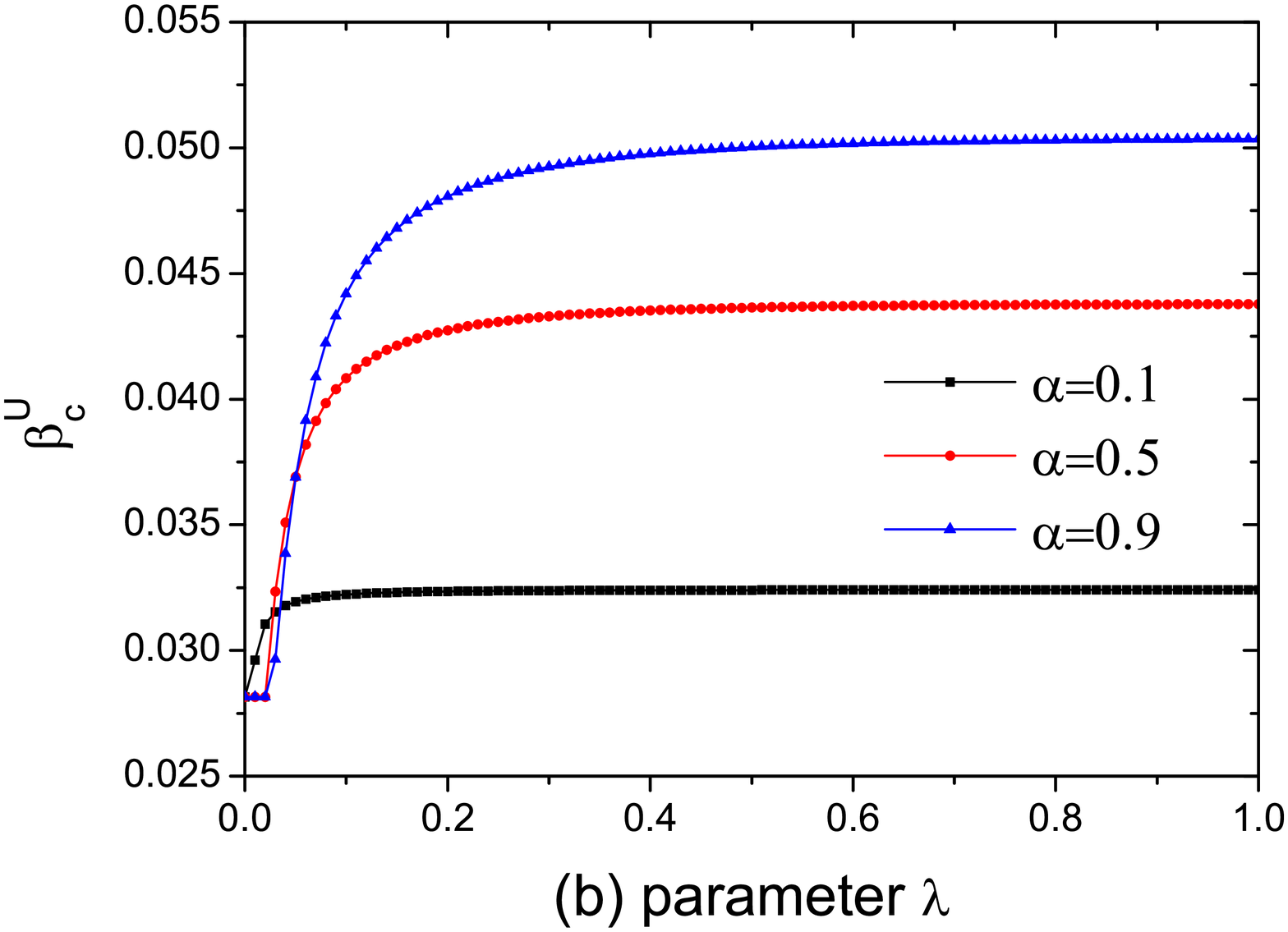}}
\end{minipage}\\[5pt] \vspace{-9pt}
\begin{minipage}[c]{.5\textwidth}
\centering \scalebox{.3}{\includegraphics{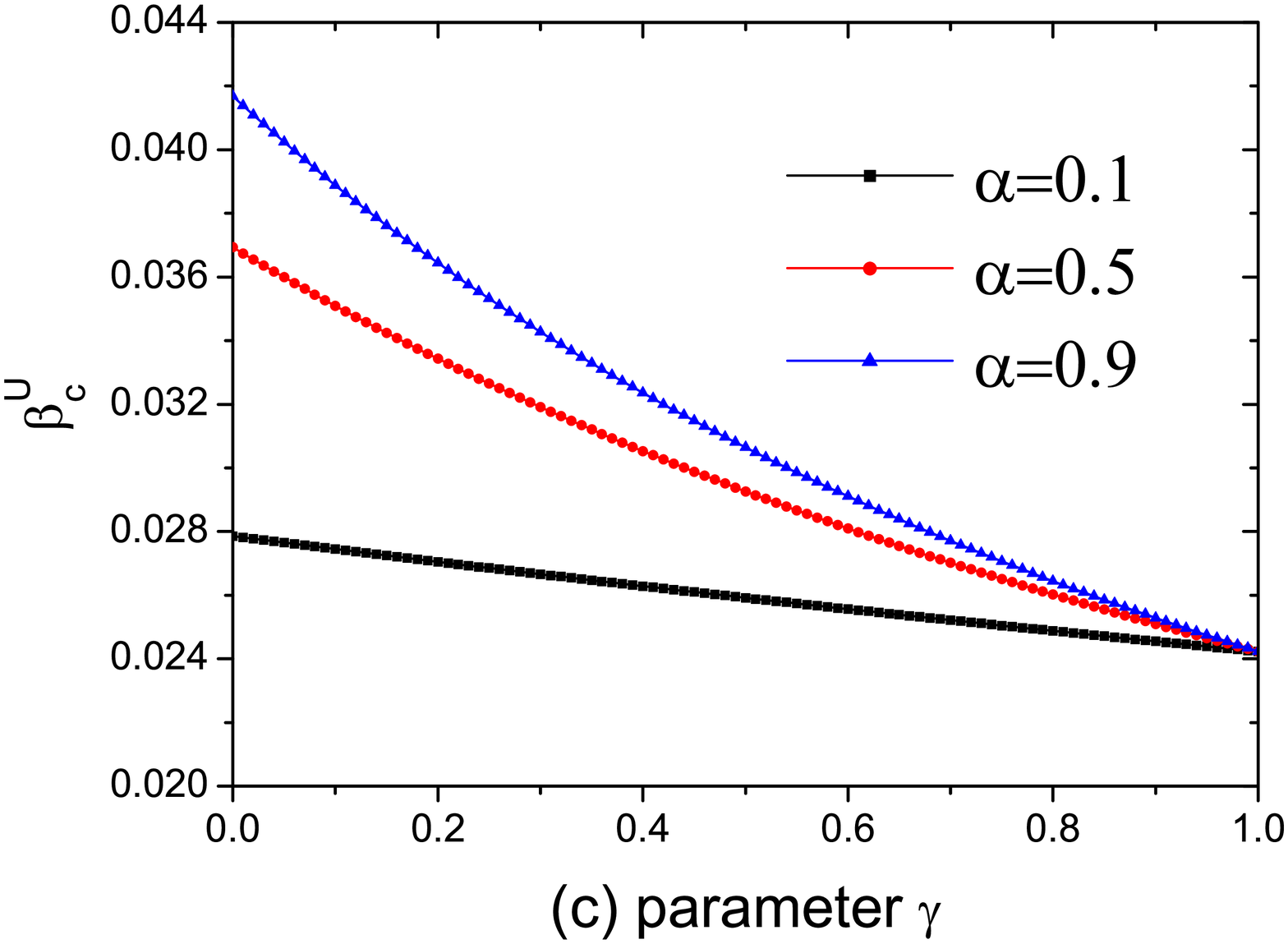}}
\end{minipage}%
\hspace{-14pt}
\begin{minipage}[c]{.5\textwidth}
\centering \scalebox{.3}{\includegraphics{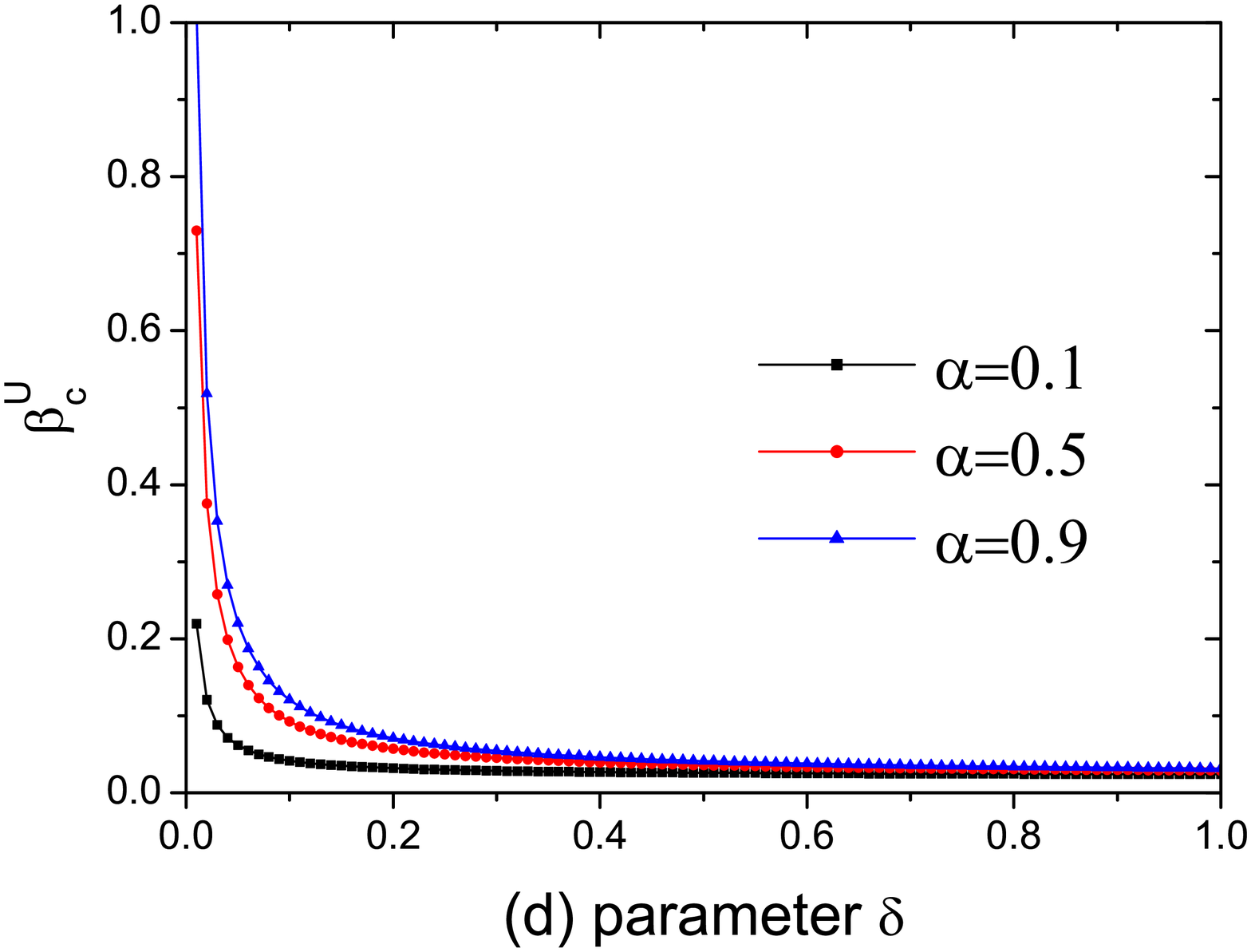}}
\end{minipage}\\[5pt]
\caption{(Color online) The epidemic threshold as a function of the model parameters $\alpha$ (a), $\lambda$ (b), $\gamma$ (c) and $\delta$ (d). In panel (a), the results are shown for two different, fixed values of $\lambda$; in panels (b)-(d), the results are shown for three distinct values of $\alpha$. In numerical simulations, unless otherwise indicated and if not varied, the remaining parameter values were always fixed to $\delta=0.6$, $\mu=0.4$, $\gamma=0$, and $\lambda=0.2$.}
\end{figure}

To this effect, let us now consider a regular multiplex network overlapped by two (random) regular networks with the identical mean degree $n_A$ for layer $A$ and $n_B$ for layer $B$. In this case, $p_i^{US}=p^{US}$ for all $i$, and $$r_i=r=\left[1-\lambda \left(1-p^{US}\right)\right]^{n_A}.$$
Then, for the matrix $H$ we have
$$
H=\left[\left(\frac{\gamma}{\delta}+\frac{1}{\alpha}\right)(1-r)+1\right]p^{US}n_B.
$$
The resulting epidemic threshold is therefore given by
 \begin{equation}
 \beta_c^U=\frac{\mu}{\left[\left(\frac{\gamma}{\delta}+\frac{1}{\alpha}\right)(1-r)+1\right]p^{US}n_B}.
 \end{equation}
Let $x=1-p^{US}$. Then $x$ satisfies the condition where
$$
\left[\left(\frac{1}{\delta}+\frac{1}{\alpha}\right)(1-(1-\lambda x)^{n_A})+1\right](1-x)=1.
$$
Furthermore, let $F(x)=\left[\left(\frac{1}{\delta}+\frac{1}{\alpha}\right)(1-(1-\lambda x)^{n_A})+1\right](1-x)$. It is then easy to show that $F(0)=1,F(1)=0$ and $F''(x)<0$. When $F'(0)\geq 1$, there exists a positive solution; otherwise, there is only one zero solution. Hence, we can find that the epidemic threshold is given by
\begin{equation}\label{RRN_threshold}
\beta_c^U=\left\{
\begin{aligned}
\frac{\mu}{\left[\left(\frac{\gamma}{\delta}+\frac{1}{\alpha}\right)(1-r)+1\right]p^{US}n_B}, \lambda>\frac{1}{\left(\frac{1}{\delta}+\frac{1}{\alpha}\right)n_A}; \\
\frac{\mu}{n_B} , \quad \quad\quad\quad\quad \lambda\leq \frac{1}{\left(\frac{1}{\delta}+\frac{1}{\alpha}\right)n_A}.
\end{aligned}
\right.
 \end{equation}
According to (\ref{RRN_threshold}), the critical threshold $\lambda_c=\frac{1}{\left(\frac{1}{\delta}+\frac{1}{\alpha}\right)n_A}$, which thus only relies on the communication layer of awareness spreading.

Besides the effects of $\alpha$ and $\lambda$, we further investigate the influence of the behavioral parameters $\gamma$ and $\delta$ on the epidemic threshold, with the main results summarized in Figure~4(c) and 4(d). Notably, we can see that in both cases, with the increasing values of these two
parameters the epidemic threshold significantly decreases. Relative to the effect of $\gamma$, the parameter $\delta$ has a much stronger effect on the epidemic threshold when $\delta$ varies in $[0,0.4]$. Hence, manipulating the information forgetting rate $\delta$ can provide a remarkably potent control strategy for epidemic spreading.

Moreover, we can see an interesting interaction effect between $\gamma$ and $\alpha$ in Figure~4(c), and another interaction effect among $\delta$ and $\alpha$ in Figure~4(d). The observed interplay between $\gamma$ and $\alpha$ indicates that unlike in the classical SIS model (when $\gamma=1$), for which the epidemic threshold is comparable across different values of $\alpha$, in our UWAU-SIS model when $\gamma<1$ and especially when the strongly aware nodes attain their full immunity (at $\gamma=0$), the epidemic threshold is elevated significantly at higher but not at low values of $\alpha$. Clearly, immunity of strongly aware nodes can play a role in epidemic threshold elevation but only if the rate $\alpha$ (at which weakly aware nodes transition to strongly aware ones) is sufficiently high. The other interplay between $\delta$ and $\alpha$ in Figure~4(d), indicates that $\alpha$ can elevate the epidemic threshold but only under minimal or fully absent information forgetting conditions; as the forgetting rate $\delta$ increases, the rate of acquisition of strong awareness $\alpha$ becomes irrelevant as to the epidemic prevalence and the epidemic threshold drastically drops down towards $\beta_c^U=0$ (Figure~4(d)).

\subsubsection{Influence of the network overlap on epidemic spreading}

In our multiplex network, there are two distinct layers which may not necessarily have the same topology or the identical set of nodes. Disease-related information diffuses over the communication layer, whereas the epidemic-causing pathogen spreads on top of 
the contact layer. Thus, a given communication network layer may not be exactly the same as the disease-spreading contact layer, but there may exist remarkable link overlaps across the communication and contact network layers \cite{WuTarik2018}.

\begin{figure}[htbp]
  \centering
\begin{minipage}[c]{.5\textwidth}
\centering \scalebox{.3}{\includegraphics{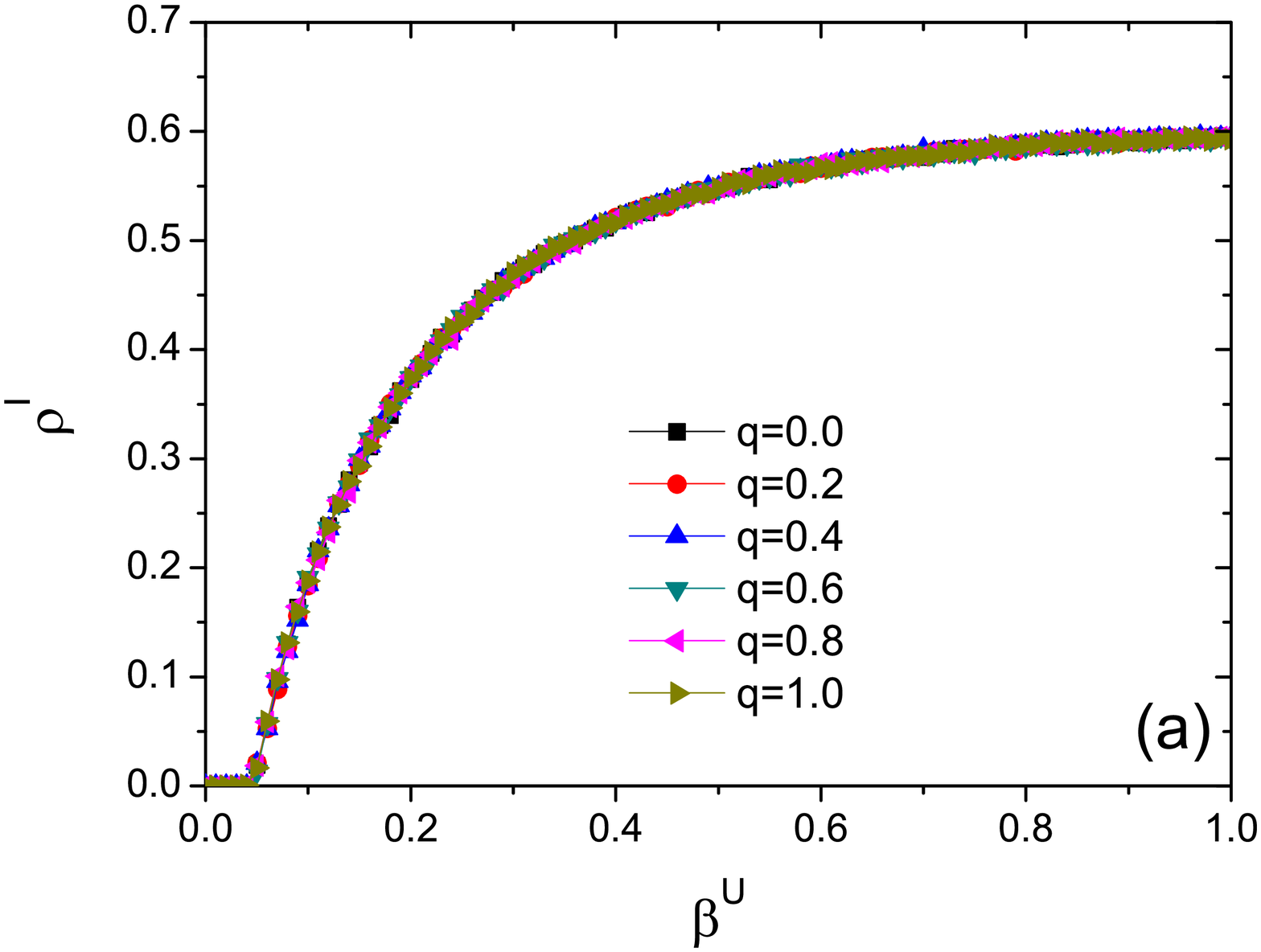}}
\end{minipage}%
\begin{minipage}[c]{.5\textwidth}
\centering \scalebox{.3}{\includegraphics{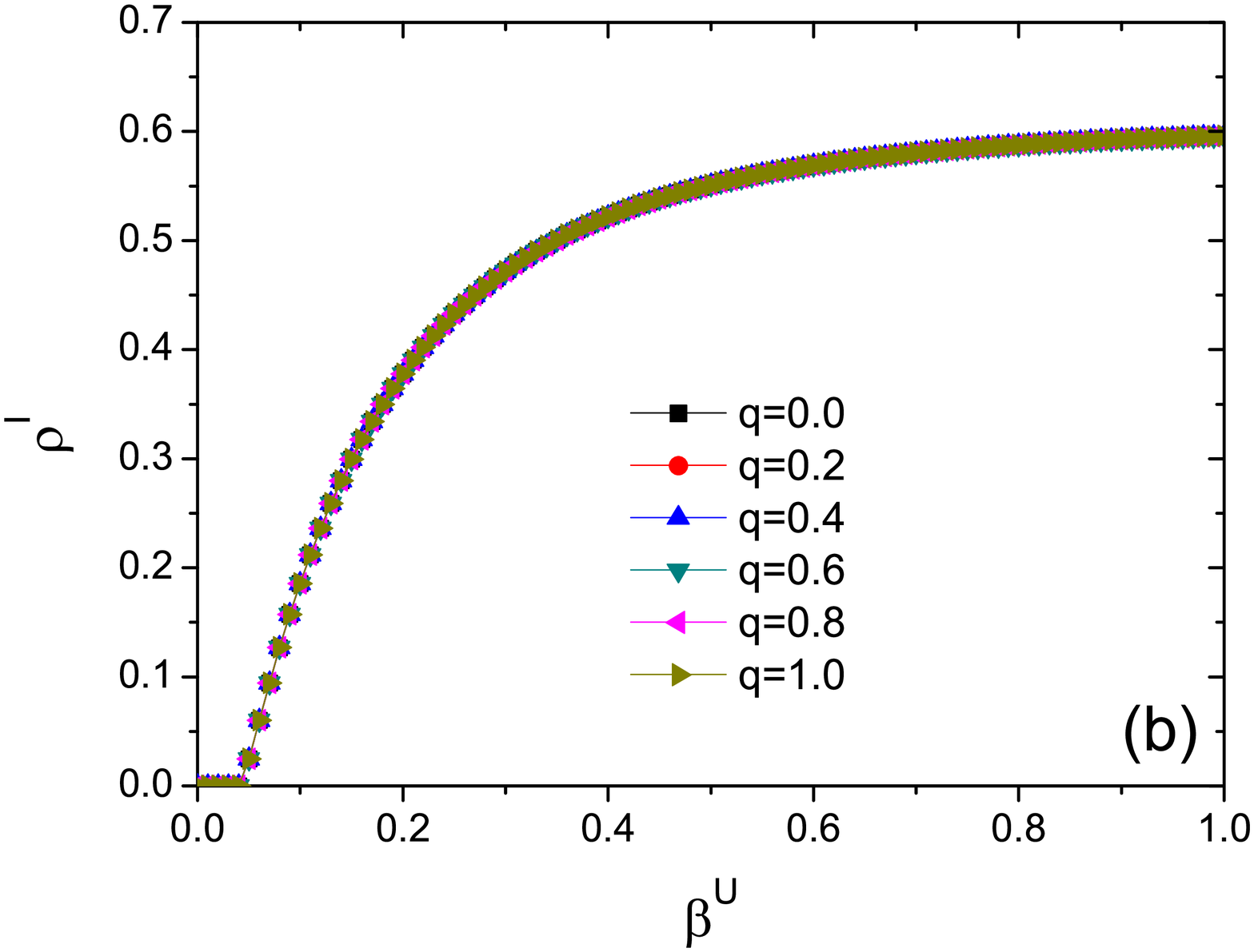}}
\end{minipage}\\[5pt]
\vspace{-10pt}
\caption{(Color online) The final epidemic size as a function of the infection rate $\beta^U$ in scale-free multiplex networks with different levels of the network links overlap $q$. Panel (a) shows the results of the stochastic simulations and panel (b) corresponds to the outcomes of the numerical simulations of the MMCA model. In all simulations, our other parameter values were selected as $\delta=0.6$, $\mu=0.4$, $\gamma=0$, $\lambda=0.5$, and $\alpha=0.5$.}
\end{figure}

We now therefore analyze the influence of network links overlap on the epidemic spreading dynamics \cite{WuTarik2018}. 
To this effect, it suffices to study the influence of the parameter $q$ on both the epidemic threshold and disease prevalence. In a series of simulation experiments, we systematically varied the values of $q$ from $q=0$ to $q=1$, and we conducted both stochastic (Figure~5(a)) and deterministic simulations (Figure~5(b)) for the same region of the parameter space. We can observe in Figure~5 that: (i) Our theoretical UWAU-SIS model agrees very well with the stochastic simulations, and quite remarkably, (ii) the parameter $q$ has negligible effect on both the epidemic threshold and the disease prevalence. 
These observations can also be understood from our theoretical findings for the special case in Eq. (\ref{RRN_threshold}), where the epidemic threshold is not related to $q$.

\section{Discussion}

Understanding the nature of the multi-leveled interplay between spreading phenomena and human behavior in complex social systems has been acknowledged as one of the main unresolved challenges of mathematical epidemiology \cite{Pastor-Satorras2015,Funketal2010,manfredi,bianconi}. Facing the uncertainties of 
the present COVID-19 pandemic, with unprecedented public health threats and catastrophic economic effects \cite{herrmann}, there is a pressing need for the development of rigorous mathematical models that would lay the foundation for a more general theory of disease-behavior dynamics, allowing us to capture the nonlinear, multi-leveled interplay between critical epidemiological and behavioral characteristics of societies affected by the pandemic crisis. 

In this paper, a novel general model of coupled disease-awareness dynamics on multiplex networks is developed. Using the multiplex network 
framework \cite{bianconi}, gradual information diffusion in the communication network layer was coupled with the endemic contagion transmission in the contact network layer, resulting in a nonlinear dynamical system in which not only disease affected the information propagation but also the disseminated information was able to influence and change the fate of a disease outbreak, thereby nontrivially reinforcing one another. 

Interestingly, we replicated a unimodal epidemic peak phenomenon occuring in the initial stages of the outbreak \cite{qcwcsf}, where the infection density jumped initially to a relatively large value but then rapidly descended to the quasi-stationary state of endemic infection (Figure 1(b) and 1(d)). Although such unimodal peak phenomena can occur in endemic transmission processes \cite{zhixiaorui}, they were largely ignored in previous coupled disease-awareness spreading models. Some eminent exceptions can be found in Ref.~\cite{Zhan2018} studying the emergence of outbreak peaks in monoplex networks during multiple epidemic waves, then in Ref.~\cite{jankochmiel} examining the influence of initially emerging opinions on infection peaks under different spreading time scales, and in Ref.~\cite{qcwcsf}, addressing simultaneous transmission of disease and information in multiplex networks. 

Our model is different from the previously widely employed probability tree based method \cite{Granell2013}, and we have demonstrated that it outperforms this competing approach both in the prediction of the time-evolution of epidemics and in the final epidemic size. In addition, we have shown in Appendix A that our UWAU-SIS model can also be formulated by using a probability tree based approach with sequential information-disease spreading, as in Ref. \cite{Granell2013}, in which disease transmission occurs after information diffusion. Moreover, for the full $\lambda-\beta^U$ phase space, we obtained a highly satisfactory agreement between Monte Carlo simulation outcomes and the numerical simulations of our MMCA based method on scale-free multiplex networks (Figure~2).

Critically, earlier studies neglected the fact that disease-related information is not disseminated instantaneously throughout the population, but that instead its diffusion progresses gradually, in stages. As a result, previous studies have failed to address potential effects of stage-based progression of awareness diffusion and its dynamic interplay with epidemic spreading in complex multilayered networks. However, as we have seen in the present study, our model revealed that future disease-behavior dynamic models should not ignore the importance of the parameter $\alpha$ (representing the transition rate between weak and strong awareness), its effect on epidemic threshold (Figure~3), and its nontrivial interaction with the rate of information spreading from aware to unaware individuals (Figure~4(a)-(b)). Specifically, we found that only informing the unaware individuals about the presence of a disease will not be helpful for preventing an epidemic outbreak unless the informed individuals can gain strong awareness of the infection risks and then promptly respond with adequate self-protective measures. These results are corroborated in part by a more recent study \cite{hongetal}, showing that time-varying self-awareness can play a role in epidemic suppression but only when sufficiently many people have gathered the necessary information and have adequately modified their behavioral responses. 

Two other interaction effects that were identified among behavioral parameters in our study revealed that the immunity of strongly aware 
nodes can influence the epidemic threshold elevation but only if $\alpha$ is sufficiently high (Figure~4(c)), and that a higher $\alpha$ can elevate the epidemic threshold but only under minimal or absent information forgetting conditions (Figure~4(d)). Specifically, with an increasing forgetting rate $\delta$, the effect of $\alpha$ becomes negligible and the epidemic threshold is consequentially drastically lowered. These findings thus indicate that manipulating the information forgetting rate $\delta$ can serve as a remarkably potent epidemic control strategy.

The observed interaction effects (Figure~4) thus evidently justify the introduction of the weakly-aware node state (W-state) and its necessary transition to the subsequent strongly-aware state (A-state) in models with coupled disease-behavior dynamics, highlighting the importance of stage-based progression models for the design of future public health policy measures. Our study thus contributes to the growing evidence showing that elevated awareness of disease risks and associated protective behavioral responses can significantly alter the trajectories of infectious diseases, further demonstrating that behavioral parameters can nontrivially interact with each other when producing their effects on the spreading dynamics of an infectious disease. Together, these findings suggest that disease-control strategies should not only generally consider the interplay between epidemic spreading and behavioral responses, but should also carefully take into account an array of complex interaction effects that exist between information-related parameters which then under special conditions exert their combined influence on disease transmission dynamics.

Different from previous studies of disease-behavior dynamics on multi-layered networks \cite{yang2016,Zhu2019}, we found that the network link overlap in our coupled dynamic system with concurrent disease-awareness spreading had no influence on the main features of an epidemic outbreak, including the epidemic threshold and infection prevalence (Figure~5). However, we note that the edge overlap may well have its important effects on the disease dynamics under other relevant model schemes \cite{WuTarik2018}. For example, when infection propagates faster than information \cite{Zhu2019}, a smaller number of overlapping links is able to effectively hinder the contagion spreading; on the contrary, when the information propagates faster than the disease, a larger amount of overlapping edges is required to effectively inhibit the epidemic outbreak. Moreover, the network links overlap does not always support the inhibitory role of information in epidemic spreading but instead it can both strengthen and weaken this inhibitory effect \cite{yang2016}, depending on the relative time-scales of the two interacting processes. Thus, it will be necessary in subsequent studies to investigate in a more detail how the influence of network overlaps on outbreak dynamics is related to the relative time-scales of interacting processes as they spread on top of multiplex networks \cite{wuinteract}. 

In addition to our main discrete-time UWAU-SIS model, we developed its corresponding continuous-time formulation (Appendix B). This development lead us to the conclusion that the epidemic threshold of our discrete-time model is lower than or equal to the one obtained from the continuous-time model version. This is an important first result that should stimulate further systematic comparisons between discrete- and continuous-time spreading processes, as they were rarely addressed in previous studies of coupled disease-awareness dynamics on multiplex networks \cite{Huang2018}, and especially because comparative analyses of this kind are lacking for most epidemic-behavior models even on monoplex networks \cite{ChenIEEE2019}.

For example, it has been recognized that discrete-time models on monoplex networks often tend to approximate real-life spreading phenomena better than their continuous-time counterparts, whereas continuous-time epidemic models that ignore both time intervals and spatial dependencies among network nodes tend to over-predict the speed of pathogen transmission \cite{ChenIEEE2019}. However, discretizing time can also significantly impair model 
performance in the case of too large state transition probabilities, inevitably causing deviations from the underlying 
continuous-time dynamics \cite{fennell2016}. A more recent extension of an EDT model for concurrent three-state spreading processes \cite{ZhouIEEE2019} suggests a promising application of a continuous-time process for modeling coupled disease-behavior dynamics in multilayered networks. Nevertheless, this model has not yet been evaluated against its discrete-time counterpart, warranting future investigation in this important direction. In particular, whilst we have shown in the present paper how to derive a continuous-time process from our discrete-time UWAU-SIS model, a challenging open question that remains to be addressed in future studies is how to actually choose between these two model types when addressing more realistic epidemic spreading scenarios.  

Remarkably, most earlier studies of coupled behavior-disease dynamics ignored temporal variability in spreading processes and were conducted usually on static networks with inactive nodes, in which the network structure remained unchanged over time. However, more recently, the MMCA method has also been applied to model more realistic scenarios of disease-awareness co-evolutionary dynamics, where both awareness diffusion and behavioral responses in multiplex networks were time-varying \cite{hongetal}, or where the information layer \cite{Guo2016} or each of the involved layers in the multiplex \cite{Yang2019} were generated by using the activity-driven modeling method \cite{jiading,Perra2012}. Future extensions of our model should therefore consider validating our curent UWAU-SIS spreading process on time-varying multilayered networks, preferentially also with a further generalization for clustered networked systems \cite{clustered}.

Promising socio-biological applications of our model, including a multi-level co-evolutionary interplay between co-infection dynamics of multiple diseases \cite{wuinteract} and competitive propagation of distinct types of information \cite{Pan2018v2} with awareness-promoting socio-behavioral processes such as cooperation \cite{yezino,hadzibeganovic4} or opinion formation \cite{penglu,Lichtenegger}, should be explored more extensively in subsequent studies. Clearly, our model is not only applicable to the study of coupled disease-behavior dynamics but it can also be generalized to address the interplay 
of awareness transmission and other contagious phenomena including spread of citations in scientific research networks \cite{plahan}, malware propagation in computer systems \cite{Rey2018}, or rumor diffusion in online social-media platforms \cite{Arruda2018}.  

\section{Conclusions}

In this paper, the interplay between gradual, stage-based progression of awareness diffusion and endemic disease transmission in multiplex networks was rigorously studied by using a probabilistic modeling approach based on the MMCA method. We analytically derived the epidemic thresholds for both discrete-time and continuous-time versions of our model, and we numerically illustrated that our model can capture the time-course and the steady state of the coupled disease-awareness dynamics. Together, these findings demonstrated that our UWAU-SIS method for coupled disease-awareness spreading was exact when the topological connectivity of each layer in the multiplex network was unclustered, and that our model outperformed a previously widely employed probability tree based approach.

More specifically, our findings revealed that informing the unaware individuals about the circulating disease is not sufficient for the prevention of an outbreak unless the distributed information triggers strong awareness of infection risks with adequate self-protective measures, and that the immunity of highly-aware individuals can substantially increase the epidemic threshold, but only if the rate of transition from weak to strong awareness is sufficiently high. Our study thus shows that awareness diffusion and other behavioral parameters can nontrivially interact when producing their effects on the spreading dynamics of an endemic disease, suggesting that future public health measures should not ignore this complex behavioral interplay and its influence on contagion transmission in multilayered networked systems.

Our method is not limited to the study of disease-awareness coevolutionary dynamics, but due to its generality, it can also be applied to a wide variety of other coupled contagious processes. Including such complex spreading phenomena as special cases in subsequent extensions of our model will be instrumental to the future development of a more general mathematical theory of coupled disease-behavior dynamics in multiplex networks. We hope that our present work will serve as a first enthusiastic step towards this formidable challenge.

\section*{Acknowledgments}
This work was supported by the National Natural Science Foundation of China under the Grant Nos. 61663015 and 62073112.

\appendix

\section{The probability tree model}

\begin{figure}[htbp]
  \centering
\begin{minipage}[c]{.45\textwidth}
\centering \scalebox{.32}{\includegraphics[17,113][552,794]{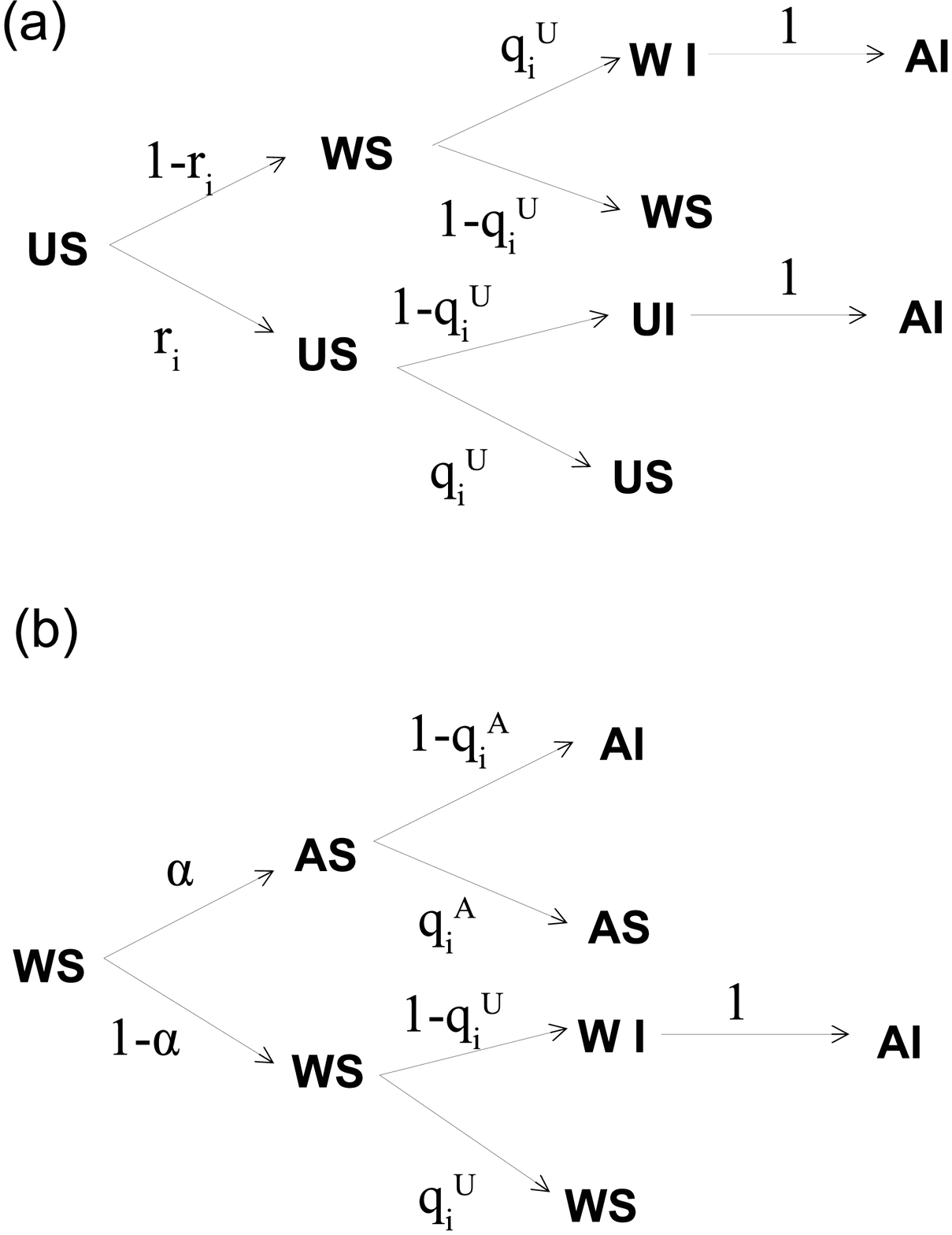}}
\end{minipage}%
\begin{minipage}[c]{.45\textwidth}
\centering \scalebox{.32}{\includegraphics[17,113][552,794]{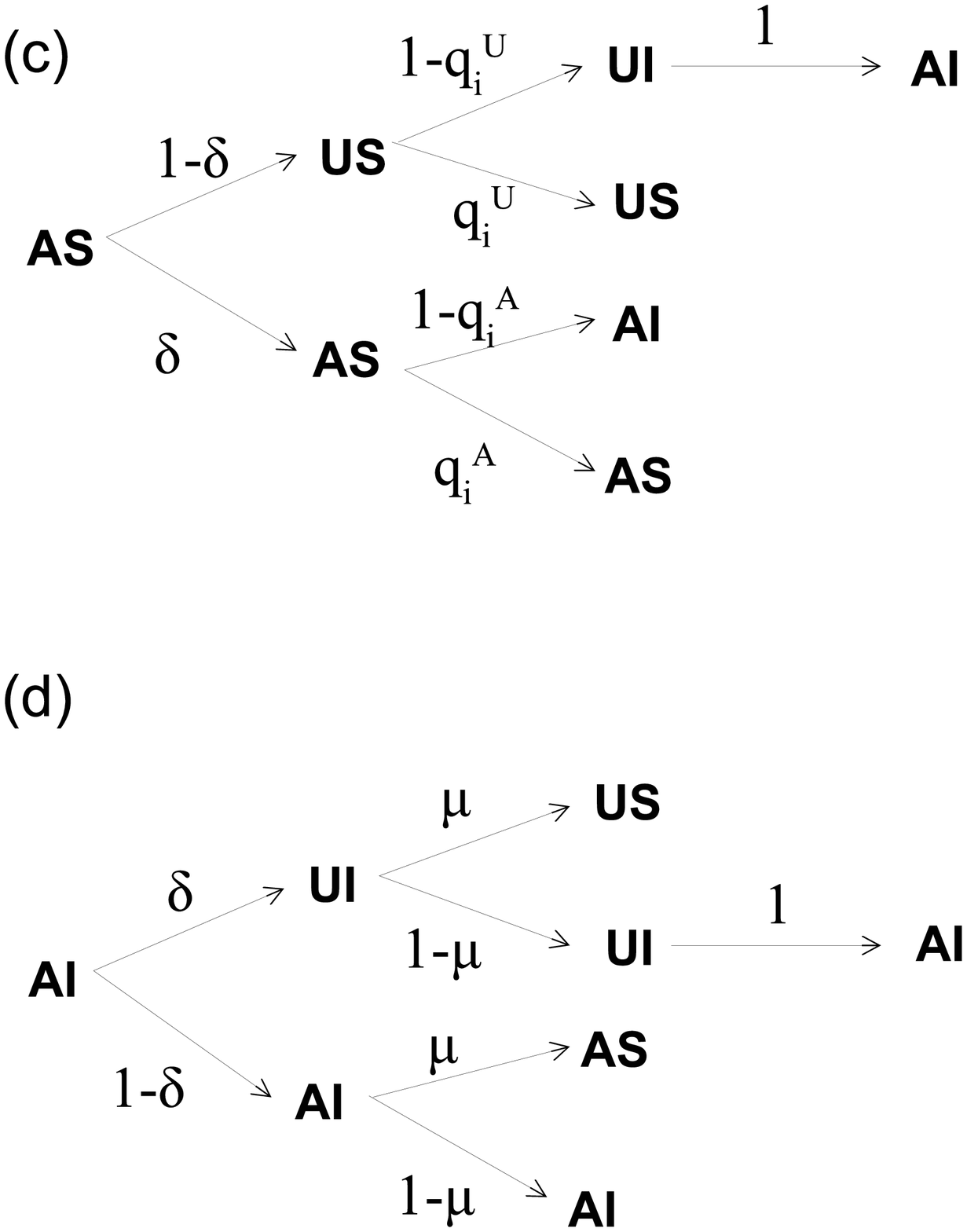}}
\end{minipage}\\[5pt]
\caption{Transition probability trees of the four possible states in the coupled UWAU-SIS spreading model: US (a), WS (b), AS (c) and AI (d). At each individual time step, the process of state-transition in the model alternates sequentially from information to disease, i.e. disease spreads after information diffusion has taken place. }
\end{figure}

The probability tree based method was employed in Ref.~\cite{Granell2013} to identify the possible state transitions and to derive the corresponding model equations for the 3-state UAU-SIS coupled disease-awareness dynamics on multiplex networks. In this method, a sequential spreading of awareness and disease processes is employed, i.e. disease transmission occurs {\it after} information diffusion. Furthermore, by following this method \cite{Granell2013}, we can also draw the probability trees for our 4-state UWAU-SIS process (as shown in Figure A.6), and the corresponding equations for our two-stage model with gradual information diffusion can then be obtained as follows:
\begin{eqnarray}\nonumber
p_i^{US}(t+1)&=&p_i^{US}(t)r_i(t)q_i^U(t)+\delta p_i^{AS}(t) q_i^U(t) +\delta \mu p_i^{AI}(t) \\\nonumber
p_i^{WS}(t+1)&=&p_i^{US}(t)[1-r_i(t)]q_i^U(t)+(1-\alpha) p_i^{WS}(t)q_i^U(t) \\\nonumber
p_i^{AS}(t+1)&=&\alpha p_i^{WS}(t) q_i^U(t)+(1-\delta)p_i^{AS}(t)q_i^A(t)+(1-\delta)\mu p_i^{AI}(t)\\\nonumber
p_i^{AI}(t+1)&=&p_i^{US}(t)[1-q_i^U(t)]+p_i^{WS}(t)\{\alpha [1-q_i^A(t)]+(1-\alpha)[1-q_i^U(t)]\}\\\nonumber
&&+p_i^{AS}(t)\{\delta [1-q_i^U(t)]+(1-\delta)[1-q_i^A(t)]\}+(1-\mu)p_i^{AI}(t).\\
\label{eqn1}
\end{eqnarray}
Similar to our previous analysis for the critical condition, it can be shown that the corresponding epidemic threshold is just equal to (\ref{threshold}). This indicates that both theoretical models ((\ref{eqn1}) and (\ref{eqnn1})) actually share exactly the same threshold condition.

\section{The continuous-time formulation of the UWAU-SIS model (\ref{eqnn1})}

In the previous parts, we derived the discrete-time disease transmission model (\ref{eqnn1}). Now, let us consider the disease transmission as a continuous-time process. To this effect, we could simply reduce the time interval $[t,t +1)$ to $[t,t+h)$ and let $h \rightarrow  0$ as in Ref. \cite{Huang2018}. In this new time interval $[t,t+h)$, the probabilities of all model parameters depend linearly on the duration of exposure \cite{Huang2018}, i.e. the model parameters are then replaced as $\delta h$, $\lambda h$, etc. For the resulting continuous-time 4-state UWAU-SIS model we then have 
\begin{eqnarray}\nonumber
p_i^{US}(t+h)&=&p_i^{US}(t)r_i(t)q_i^U(t)+\delta h p_i^{A_1S}(t) q_i^A(t) +\delta \mu h^2 p_i^{A_1I}(t) \\\nonumber
      &=:&p_i^{US}(t){\rm Prob}_h(US\rightarrow US)+p_i^{A_1S}(t){\rm Prob}_h(AS\rightarrow US)+\delta \mu h^2 p_i^{A_1I}(t) \\\nonumber
p_i^{WS}(t+h)&=&p_i^{US}(t)[1-r_i(t)]q_i^U(t)+(1-\alpha h) p_i^{WS}(t)q_i^U(t) \\\nonumber
&=:&p_i^{US}(t){\rm Prob}_h(US\rightarrow WS)+p_i^{WS}(t){\rm Prob}_h(WS\rightarrow WS)\\\nonumber
p_i^{AS}(t+h)&=&\alpha h p_i^{WS}(t) q_i^U(t)+(1-\delta h)p_i^{AS}(t)q_i^A(t)+(1-\delta h)\mu h p_i^{AI}(t)\\\nonumber
&=:&p_i^{WS}(t){\rm Prob}_h(WS\rightarrow AS)+p_i^{AS}(t){\rm Prob}_h(AS\rightarrow AS)+(1-\delta h) \mu h p_i^{AI}(t) \\\nonumber
p_i^{AI}(t+h)&=&p_i^{US}(t)[1-q_i^U(t)]+p_i^{WS}(t)[1-q_i^U(t)]+p_i^{AS}(t)[1-q_i^A(t)]+(1-\mu h)p_i^{AI}(t)\\\nonumber
&=:&p_i^{US}(t){\rm Prob}_h(US\rightarrow AI)+p_i^{WS}(t){\rm Prob}_h(WS\rightarrow AI)\\\label{eqnnv1}
&&+p_i^{AS}(t){\rm Prob}_h(AS\rightarrow AI)+(1-\mu h) p_i^{AI}(t).
\end{eqnarray}
Furthermore, for an adequately small $h$, we obtain that
\begin{eqnarray}\nonumber
&&{\rm Prob}_h(US\rightarrow US)\\\nonumber &=&r_i(t)q_i^U(t)\\\nonumber
&=& \left[1-\lambda h\sum_{j=1}^N a_{ij}(p_{j}^{WS}(t)+p_{j}^{AS}(t)+p_{j}^{AI}(t))+O(h^2)\right]\left[1-\beta^U h\sum_{j=1}^N b_{ij}p_{j}^{AI}(t)+O(h^2)\right]\\\label{v0}
&=&1-\lambda h\sum_{j=1}^N a_{ij}(p_{j}^{WS}(t)+p_{j}^{AS}(t)+p_{j}^{AI}(t))-\beta^U h\sum_{j=1}^N b_{ij}p_{j}^{AI}(t)+O(h^2).
\end{eqnarray}
Analogously,
\begin{eqnarray}\label{v1}
{\rm Prob}_h(AS\rightarrow US)=\delta h q_i^A(t)=\delta h +O(h^2).
\end{eqnarray}
\begin{eqnarray}\nonumber
&&{\rm Prob}_h(US\rightarrow WS)\\\nonumber
&=&[1-r_i(t)]q_i^U(t)\\\nonumber
&=& \left[\lambda h\sum_{j=1}^N a_{ij}(p_{j}^{WS}(t)+p_{j}^{AS}(t)+p_{j}^{AI}(t))+O(h^2)\right]\left[1-\beta^U h\sum_{j=1}^N b_{ij}p_{j}^{AI}(t)+O(h^2)\right]\\\label{v2}
&=&\lambda h\sum_{j=1}^N a_{ij} (p_{j}^{WS}(t)+p_{j}^{AS}(t)+p_{j}^{AI}(t))+O(h^2).
\end{eqnarray}
\begin{eqnarray}\nonumber
&&{\rm Prob}_h(WS\rightarrow WS)\\\nonumber
&=&(1-\alpha h)q_i^U(t)\\\nonumber
&=&(1-\alpha h)\left[1-\beta^U h\sum_{j=1}^N b_{ij}p_{j}^{AI}(t)+O(h^2)\right]\\\label{v3}
&=&1-\left[\alpha+\beta^U \sum_{j=1}^N b_{ij} p_{j}^{AI}(t)\right] h+O(h^2).
\end{eqnarray}

\begin{eqnarray}\nonumber
&&{\rm Prob}_h(WS\rightarrow AS)\\\nonumber
&=&\alpha h q_i^U(t)\\\nonumber
&=& \alpha h\left[1-\beta^U h\sum_{j=1}^N b_{ij}p_{j}^{AI}(t)+O(h^2)\right]\\\label{v4}
&=&\alpha h+O(h^2).
\end{eqnarray}
\begin{eqnarray}\nonumber
&&{\rm Prob}_h(AS\rightarrow AS)\\\nonumber
&=&(1-\delta h)q_i^A(t)\\\nonumber
&=& (1-\delta h)\left[1-\beta^A h\sum_{j=1}^N b_{ij} p_{j}^{AI}(t)+O(h^2)\right]\\\label{v5}
&=&1-\left[\delta+\beta^A \sum_{j=1}^N b_{ij} p_{j}^{AI}(t)\right]h+O(h^2).
\end{eqnarray}
\begin{eqnarray}\nonumber
&&{\rm Prob}_h(US\rightarrow AI)\\\nonumber
&=&{\rm Prob}_h(WS\rightarrow AI)=1-q_i^U(t)\\\label{v6}
&=& \beta^U h\sum_{j=1}^N b_{ij} p_{j}^{AI}(t)+O(h^2).
\end{eqnarray}
\begin{eqnarray}\label{v7}
{\rm Prob}_h(AS\rightarrow AI)=1-q_i^A(t)=\beta^Ah\sum_{j=1}^N b_{ij} p_{j}^{AI}(t)+O(h^2).
\end{eqnarray}
Finally, upon substituting Eqs.(\ref{v0})-(\ref{v7}) in (\ref{eqnnv1}), dividing by $h$, as well as letting $h\rightarrow 0$, we can obtain 
the continuous-time model equations as follows
\begin{eqnarray}\nonumber
\dot{p}_i^{US}&=&-p_i^{US}\left[\lambda \sum_{j=1}^N a_{ij}(p_{j}^{WS}+p_{j}^{AS}+p_{j}^{AI})+\beta^U \sum_{j=1}^N b_{ij}p_{j}^{AI}\right]+\delta p_i^{AS} \\\nonumber
\dot{p}_i^{WS}&=&\lambda p_i^{US} \sum_{j=1}^N a_{ij}(p_{j}^{WS}+p_{j}^{AS}+p_{j}^{AI})- p_i^{WS}\left[\alpha+\beta^U \sum_{j=1}^N b_{ij} p_{j}^{AI}\right] \\\nonumber
\dot{p}_i^{AS}&=&\alpha p_i^{WS}-p_i^{AS}\left[\delta+\beta^A \sum_{j=1}^N b_{ij}p_{j}^{AI}\right]+\mu p_i^{AI}\\\label{eqnnv2}
\dot{p}_i^{AI}&=&\beta^U \left[p_i^{US}\sum_{j=1}^N b_{ij} p_{j}^{AI}+p_i^{WS}\sum_{j=1}^N b_{ij} p_{j}^{AI}\right]+\beta^Ap_i^{AS}\sum_{j=1}^N b_{ij} p_{j}^{AI}-\mu p_i^{AI}.
\end{eqnarray}
In order to determine the epidemic threshold of the system (\ref{eqnnv2}), we can write a linear form of the UWAU-SIS model around the fixed points $p_i^{US}(t)=p_i^{US},p_i^{WS}(t)=p_i^{WS},p_i^{AS}(t)=p_i^{AS},p_i^{AI}(t)=0$. Firstly, from Eq.~(\ref{eqnnv2}), one finds that the equilibrium states satisfy the following relations:
\begin{eqnarray}\label{relationv1}
\lambda p_i^{US}\sum_{j=1}^N a_{ij}(p_{j}^{WS}+p_{j}^{AS})=\delta p_i^{AS}=\alpha p_i^{WS},
\end{eqnarray}
and
\begin{eqnarray}\label{relationv2}
p_i^{US}+p_{j}^{WS}+p_{j}^{AS}=1.
\end{eqnarray}
By further employing Eqs.(\ref{relationv1}) and (\ref{relationv2}), the linear system of the infectious compartment in Eq.~(\ref{eqnnv2}) is given by
\begin{eqnarray}\label{linv1}
\dot{p}_i^{AI}&=&-\mu p_i^{AI}+[\beta^Ap_i^{AS}+\beta^U( p_i^{US}+ p_i^{WS})]\sum_{j=1}^N b_{ij} p_{j}^{AI}.
\end{eqnarray}
Thus, in accordance with the next generation matrix method \cite{van2002}, the epidemic threshold $\beta_c^U$ satisfies
\begin{equation}\label{threshold-f}
 \beta_c^U=\frac{\mu}{\Lambda_{\rm max}(L)}.
 \end{equation}
From Eq.~(\ref{jacobin}), one can further find that the matrix $H$ can be rewritten as
\begin{eqnarray}\label{jacobin-f}
L_{ij}=\left[1-\frac{\alpha(1-\gamma)}{\alpha+\delta}\left(1-p_i^{US}\right)\right]b_{ij},
\end{eqnarray}
where $p_i^{US}$ obeys
$$
\left[\left(\frac{\lambda}{\delta}+\frac{\lambda}{\alpha}\right)\sum_{j=1}^Na_{ij}(1-p_j^{US})+1\right]p_i^{US}=1.
$$
By Bernoulli's inequality, we then have
$$
\prod_{j\in \Gamma^a_{i}}{\left[1-\lambda \left(1-p_{j}^{US}\right)\right]}\geq 1-\lambda \sum_{j=1}^Na_{ij}(1-p_j^{US}).
$$
From this, we finally have that 
$$
\Lambda_{\rm max}(L)\leq \Lambda_{\rm max}(H).
$$ 
Thus, the epidemic threshold of the discrete-time UWAU-SIS model is not larger than that obtained from its 
continuous-time formulation.
%

\bibliographystyle{elsarticle-num}

\end{document}